\documentclass[final,1p]{elsarticle}

\biboptions{sort&compress}
\DeclareGraphicsExtensions{.pdf,.png,.jpg}

\usepackage{amssymb}
\usepackage{amsmath}
\usepackage{bm}

\ExplSyntaxOn
\cs_gset:Npn \__first_footerline:
  { \group_begin: \small \sffamily \__short_authors: \group_end: }
\ExplSyntaxOff

\begin{document}

\begin{frontmatter}




\title{Comparison of Bending-Energy Discretization Methods for Anisotropic Meshes in Morphogenetic Simulations}

\author[1]{Tomohiro Mimura\corref{cor1}}
\ead{mimura.tomohiro.n28@kyoto-u.jp}

\cortext[cor1]{Corresponding author}
\author[2]{Yasuhiro Inoue}
\ead{inoue.yasuhiro.4n@kyoto-u.ac.jp}
\address{
  Department of Micro Engineering, Graduate School of Engineering, 
  Kyoto University, Kyoto 615-8540, Japan}

\begin{abstract}
    Accurately modeling bending energy in morphogenetic simulations is crucial, especially when dealing with anisotropic meshes where remeshing is infeasible due to the biologically meaningful entities of vertex positions (e.g., cells). This study addresses the underexplored question of which bending-energy discretization methods are most accurate and suitable for such simulations.

    The evaluation consists of two stages: First, the accuracy of each method is tested by comparing predicted bending energy and force against theoretical values for two benchmark cases--a wrinkled planar sheet and a smooth spherical sheet. Second, we simulate the formation of wrinkles in a planar sheet caused by anisotropic cell division, analyzing the resulting wavenumber patterns for two division orientations: uniaxial and random.
    
    The results highlight that the choice of the optimal discretization method depends on the application. For simulations requiring precise quantitative predictions, the Hamann model demonstrates superior accuracy. Conversely, for simulations where qualitative trends in morphology are of primary interest, the J\"{u}licher model provides a computationally efficient alternative. These findings provide guidance for selecting appropriate bending-energy discretization methods in morphogenetic simulations, ultimately leading to more accurate and efficient modeling of complex biological forms.
\end{abstract}


\begin{keyword}
anisotropic mesh \sep bending-energy \sep discretization method \sep morphogenesis \sep cell-center model

\end{keyword}

\end{frontmatter}



\section{Introduction}\label{sec1}
Thin shell deformation has been applied to solve various problems across diverse fields, including the study of vesicle and red blood cell (RBC) shapes in fluids \cite{Kraus1996,LAC2004,Noguchi2005,Misbah2006,ABREU2014}, buckling of thin films placed on fluid surfaces \cite{Pocivavsek2008,Diamant2011,Audoly2011,Demery2014,Oshri2015}, and cloth simulations \cite{volino1995,Bridson2003, Grinspun2003, Pabst2008, ZHOU2008}. Numerical solution methods are indispensable for analyzing these complex phenomena. A widely used method involves representing continuous surfaces through triangular element decomposition. To address mechanical problems using this discrete framework, it is necessary to account for both in-plane elastic and bending deformations through a suitable bending-energy discretization method.

Discretization of bending energy is based on Canham's formula using principal curvatures \cite{CANHAM197061} and Helfrich's formula using mean and Gaussian curvatures \cite{Helfrich1973}. However, these methods require the second- and fourth-order derivatives of surface coordinates to calculate the bending energy and resulting force, respectively \cite{Guckenberger2016,Koumoutsakos2020}. Various methods exist to approximate these parameters. For example, Kantor and Nelson used the angle between two adjacent triangles to express the bending energy and simulated the phase transition of a shape in polymer films \cite{Kantor1987}. Later, J\"{u}licher \cite{Julicher1996} and Gommper and Kroll \cite{Gompper1996} constructed models to represent the discrete curvatures of vesicles. Since then, various discretization methods have been developed, and compared \cite{Guckenberger2016,Koumoutsakos2020,Tsubota2014,Hoore2018}. Guckenberger et al. \cite{Guckenberger2016} calculated the mean curvature and force of red blood cells, evaluated the errors against theoretical values, and notes that the errors increase for non-uniform triangular meshes. Bian et al. \cite{Koumoutsakos2020} compared equilibrium shapes calculations for lipid bilayers and highlights that different discretization methods produce varying results. Therefore, systems that can be represented by isotropic triangular meshes, such as vesicles and RBCs, have been extensively studied.

However, not all phenomena can be accurately represented using isotropic triangular meshes. Biological processes often involve complex, dynamic shapes that require anisotropic modeling approaches. In biology, the cell-center model \cite{lattice-free-model2001,Pathmanathan2009,Osborne2010,Li2014,Lomas2014,Soumya2015,SPV2016,ActiveVM2017,Bonilla2020} is widely employed to study cell movement and tissue formation by representing individual cells as vertices within a mesh. This approach allows simulating diverse processes, such as cell movement in intestinal crypts \cite{lattice-free-model2001,Osborne2010} and cell migration \cite{Li2014,Soumya2015,SPV2016,ActiveVM2017,Bonilla2020} in two dimensions, by defining cell--cell interactions based on vertex connectivity. We have previously extended this model to 3D to simulate the morphogenesis of sheet-like epithelial tissues \cite{MIMURA2023}. This model allowed us to explore how cells utilize mechanisms, such as cell division \cite{Guillot2013,GODARD2019114} and apical constriction \cite{ApicalConstriction,SAWYER20105}, to control global tissue deformation and buckling \cite{Nelson2016, LEMKE2021}.

Cell-division axis orientation plays a crucial role in shaping the tissue morphology during morphogenesis \cite{GILLIES2011R599, Adachi2018, TANG2018313}. However, this process often leads to anisotropic cell arrangements and, consequently, anisotropic meshes. Because the vertices in these meshes represent individual cells, remeshing to maintain isotropy during the simulations is not feasible. Therefore, research on the accuracy of different-bending energy discretization methods for anisotropic meshes is limited.

This study addresses this gap by comprehensively evaluating the accuracy and applicability of various bending-energy discretization methods for simulating anisotropic meshes using a cell-center model. Specifically, we compared 13 existing discretization methods by incorporating them into the cell-center model and simulating the 3D deformation of the monolayer epithelial sheet tissue. Two sets of simulations were conducted to assess the accuracy of each method. First, to investigate the bending-energy error in anisotropic meshes, we prepared anisotropic sheets with simple wrinkle patterns, allowing for theoretical bending-energy calculations, and compared the accuracies of each method across different cell configurations. Next, we compared them under the dynamic conditions by simulating sheet deformation owing to cell division with two axis orientations: uniaxial, which leads to anisotropic meshes, and random, which maintains mesh isotropy. We simulated epithelial folding caused by cell proliferation to identify the most suitable bending-energy discretization method for accurately capturing morphogenetic events in the cell-center model.

\section{Cell-center model}\label{sec2-1}
\begin{figure}[p]
    \includegraphics[scale=0.22]{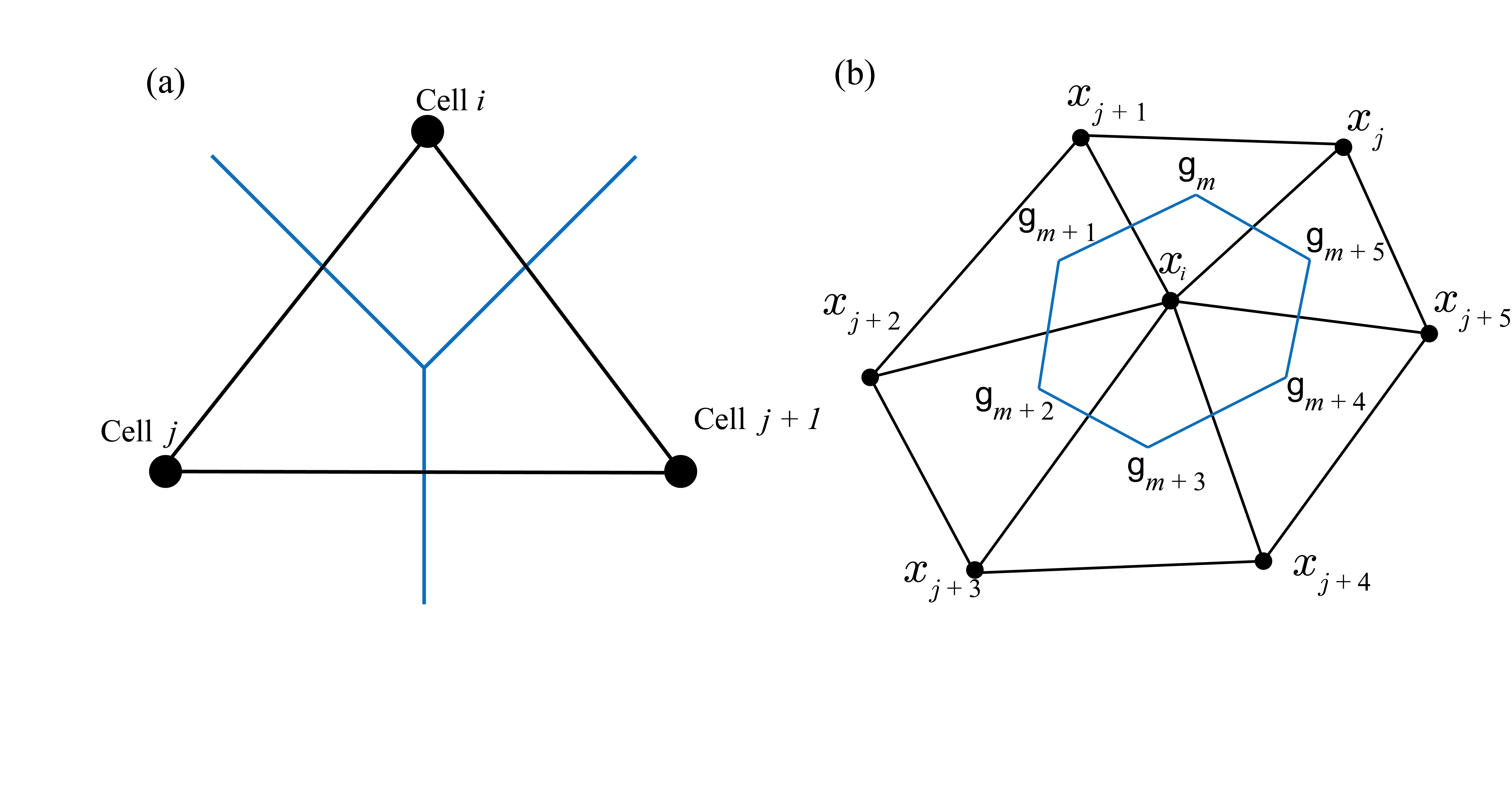}
    \caption{(a) Triangular element: the blue boundaries represent cell boundaries, dots represent cell center, and black lines connect adjacent cell centers.
    (b) Cell shape $i$: Let $\bm{x}_i$ be the position vector of cell $i$. Its shape is defined using multiple triangular elements. Specifically, by sequentially connecting the centers of mass $\bm{g}_m$ of these triangles from $\bm{g}_m$ to $\bm{g}_{m+5}$, a polygon that approximates the shape of cell $i$ is formed.
    }
    \label{Fig.2-1}
\end{figure}

In this study, the cell-center model was used to construct a triangular mesh by connecting adjacent cell centers in the epithelial sheet. As illustrated in Fig. \ref{Fig.2-1}(a), triangular elements are formed by connecting the centers of adjacent cells, assuming three-cell adjacency. The cell shape is then defined by connecting the centers of mass of the triangles that share a common cell, as shown in Fig. \ref{Fig.2-1}(b). This approach eliminates the need to explicitly maintain the Delaunay property and simplifies mesh construction. The model simulates the tissue dynamics by performing mechanical and geometrical calculations at each time step.

\subsection{Mechanical calculation}
The cell-center coordinates $\bm{x_i}$ evolve over time based on the equations of motion at the cell center $i$.
\begin{equation}\label{Eq.1}
    \eta\frac{d\bm{x_i}}{dt}=-\nabla _{\bm{x_i}}E+\bm{F^B_i}  
\end{equation}
where $\eta$ is the viscosity coefficient, $E$ is the energy function of the system excluding the bending stiffness, and $\bm{F^B_i}$ is the force applied to cell $i$ by the bending energy. A detailed explanation is provided in Section 3.

The energy function $E$ comprises the mechanical energy $E_{\rm{Mech}}$ excluding the bending stiffness and the constraint energy $E_{\rm{Center}}$ to satisfy the cell-center constraint \cite{MIMURA2023}.
\begin{equation}\label{Eq.2}
    E=E_{\rm{Mech}}+E_{\rm{Center}}
\end{equation}

The mechanical energy $E_{\rm{Mech}}$ is composed of the edge elastic energy between adjacent cells $E_L$, area elastic energy of the triangle connecting adjacent cells $E_S$, excluded volume energy of the cell $E_{\rm{Rep}}$, and the constraint energy from the surrounding environment when deformed out-of-plane $E_Z $.

\begin{equation}\label{Eq.3}
    E_{\rm{Mech}}=E_L+E_S+E_{\rm{Rep}}+E_Z
\end{equation}

The edge elastic energy $E_L$ is the sum of the energies required to adjust the edge length $L_{ij}$ between adjacent cells $i$ and $j$ to the equilibrium length $L_{\rm{eq}}$. This relationship is expressed using the proportionality constant $K_L$ as follows:
\begin{equation}\label{Eq.4}
    E_L=\sum_{<i,j>}^{}\frac{1}{2} K_{L}(L_{ij}-L_{\rm{eq}})^{2}
\end{equation}
where $<i,j>$ denotes the sum of all edges between two adjacent cells.

The area elastic energy of a triangle $E_S$ is the sum of energies that maintains its area $S_{ijk}$, created by cells $i$,$j$,$k$, at the equilibrium area $S_{\rm{eq}}$ using the area elastic constant $K_S$ as follows:
\begin{equation}\label{Eq.5}
    E_S=\sum_{<i,j,k>}^{}\frac{1}{2}K_{S}(S_{ijk}-S_{\rm{eq}})^{2}
\end{equation}
where $<i,j,k>$ indicates that the sum is obtained over all triangles composed of three adjacent cells.

The excluded volume energy of cell $E_{\rm{Rep}}$ is the sum of the energies that induce a repulsive force repulsion when the distance $r_{ij}$ between cells $i$ and $j$ is smaller than the threshold $r_c$, with $K_{\rm{Rep}}$ as a proportionality constant.
\begin{equation}\label{Eq.6}
    E_{\rm{Rep}}=\left\{
      \begin{array}{l}
      \sum_{i}\sum_{j<i}\frac{1}{2}K_{\rm{Rep}}(\frac{r_{ij}}{r_c}-1)^{2} \quad(r_{ij}<r_c)\\
      0 \quad (r_{ij}\geq r_c)
      \end{array}
    \right.
\end{equation}

The out-of-plane deformation constraint energy $E_Z$ is constrains the displacement of the $z$ coordinate of cell $i$ from its initial position to zero. It is weighted by the proportionality constant $K_Z$ and the area $A_i$ occupied by cell $i$ as follows:
\begin{equation} 
    A_i=\frac{1}{3}\sum_{<i,j,k>}^{}\bm{S}_{ijk}\cdot \bm{n}_i.
\end{equation}
where $\bm{S}_{ijk}$ is the area vector of triangle $ijk$ with a vertex at cell $i$.

The normal vector $\bm{n}_i$ of cell $i$ is obtained by linear summation of the weighting factor $w_t$ and unit normal vector $\bm{u}^t$ for all triangles $t$ sharing cell $i$.
\begin{equation}\label{Eq.NormalVector}
    \bm{n}_i=\frac{\sum_{<i,j,k>}^{}w_t \bm{u}^t}{\|\sum_{<i,j,k>}^{}w_t \bm{u}^t\|}    
\end{equation}
where $\|\cdots\|$ denotes the $L^2$-norm of a vector. 

Based on a previous study \cite{Max2000}, the weighting factor $w_t$ is calculated by dividing the area of each triangle by the square of the products of the lengths of its two sides adjacent to cell $i$. In this study, we employed this relationship even when vertex normal vectors were used for calculating the bending model.

\begin{equation}\label{Eq.7}
  E_Z=\sum_{i}\frac{1}{2}K_ZA_id_{iz}^2
\end{equation}
where $d_{iz}$ denotes the displacement of the $z$ coordinate of cell $i$ from its initial position.

The cell-center constraint energy $E_{\rm{Center}}$ forces the model to satisfy the constraint that the cell center should coincide with the center of mass of the polygon representing its 2D shape \cite{MIMURA2023}.
\begin{equation}\label{ecenter}
    E_{\rm{Center}}=\frac{1}{2}K_{\rm{Center}}\sum_{i}^{}\|(\bm{x}_i-\bm{x}_i^{\rm{Center}})-[(\bm{x}_i-\bm{x}_i^{\rm{Center}})\cdot \bm{n}_i]\bm{n}_i\|^{2}
\end{equation}
\begin{equation}
    \bm{x}_i^{\rm{Center}}=\frac{1}{N_i}\sum_{<i,j>}^{}\bm{x}_j
\end{equation}

where $\bm{x}_i^{\rm{Center}}$ is the polygonal center-of-mass vector representing the 2D shape of cell $i$.
The number of cells connected to cell $i$ is $N_i=\sum_{<i,j>}^{}1$.

\subsection{Geometric calculation}
In this study, we investigated cell intercalation, wherein the adjacency between cells changes without altering the number of cells, and cell division, which increases the number of cells.

\begin{figure}[p]
    \centering
    \includegraphics[scale=0.55]{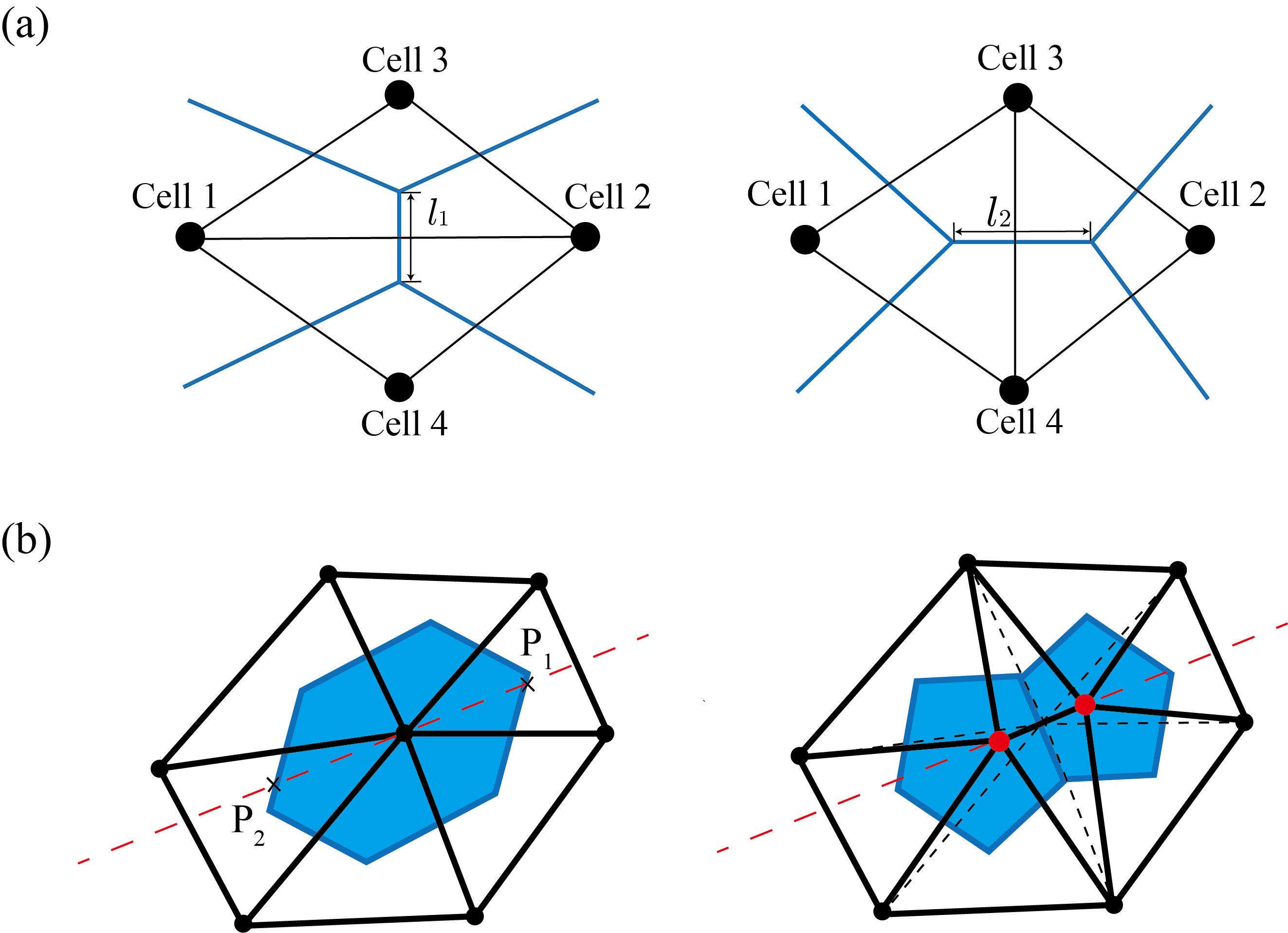}
    \caption{(a) Cell intercalation: Cells 3 and 4 interrupt Cells 1 and 2, and the adhesive relationship is updated accordingly. The blue lines represent the cell contours, whereas the black lines represent the cell adjacency. This operation is performed only when the distance $l_1$ between the mass centers of the two triangles before the flip is less than the length threshold $l_{th}$ and $l_1$ before the flip is less than that ($l_2$) after the flip. (b) Cell division: (left) Finding the two intersection points, $P_1$ and $P_2$, between the projected polygon and the division axis; (right) Setting the coordinates of the two cells after cell division such that the line segment $P_1$-$P_2$ is internally divided in a 1:1:1 ratio. The dotted red line represents the axis of division. Cell $i$ divides into two daughter cells, and the red dots denote the newly formed cells. }
    \label{Fig.2-2}
\end{figure}

Cell intercalation, wherein the cell arrangement changes while modeling the adhesion between neighboring cells, was represented using a flip operation. As shown in Fig. \ref{Fig.2-2}(a), this operation swaps the diagonals of a quadrangle formed by two adjacent triangles across a line.
This operation is performed only when the distance $l_1$ between the mass centers of the two triangles before the flip is less than the length threshold $l_{th}$ and $l_1$ before the flip is less than that ($l_2$) after the flip. To prevent continuous flips of the same edge, no flips occur on the same edge during a cool-down time of $\tau_{ct}$ after the flip.

Cell division is represented as follows: When a cell divides, the surrounding points are projected onto the plane defined by the normal vector $\bm{n}_i$ of the dividing cell $i$. The intersection of the vector passing through vertex $i$, specifying the division direction, and the polygon representing the cell shape is determined (Fig. \ref{Fig.2-2}(b)left). The coordinates of the two daughter cells are set such that the line segment formed by the two intersections is divided in a 1:1:1 ratio (Fig. \ref{Fig.2-2}(b)right). The adjacency relationships are updated and the cells are reconnected to maintain the triangular mesh.

Unlike previous studies \cite{MIMURA2023, okuda2013-b, Inoue2017, Inoue2020}, we eliminated randomness by assuming that cell $i$ divides when its cell time $\tau_{i}$ reaches $\tau_{i}^{\rm{cycle}}$. The initial cell time was sampled uniformly from zero to $\tau_{i}^{\rm{cycle}}$.

\section{Bending-energy model}\label{sec2-2}
According to Helfrich \cite{Helfrich1973}, the free energy of curvature $E_B$ can be expressed as follows:
\begin{equation}\label{Eq.9}
    E_B=2K_B\int_{}^{}(H-H_0)^2dA+K_G\int_{}^{}GdA.
\end{equation}
where $H$, $H_0$, and $G$ denotes the mean, spontaneous, and Gaussian curvatures, respectively, and $K_B$ and $K_G$ are the proportionality constants for the mean and Gaussian curvature terms, respectively. Under the condition of an invariant topology, the second term can be neglected because it remains constant according to the Gauss--Bonnet theorem.

By defining the discretized energy $E'_B$, the force $\bm{F^B_i}$ owing to the bending stiffness applied to cell $i$ can be written as follows:
\begin{equation}\label{Eq.force1}
    \bm{F^B_i}=-\nabla_{\bm{x_i}}E'_B
\end{equation}

Discrete expressions for $\bm{F^B_i}$ can be broadly classified into two methods: one using the dihedral angles of triangles and the other using the discretized mean curvature. The models considered in this study for each method are introduced in Sections \ref{sec2-3-1} and \ref{sec2-3-2}.

\subsection{Method using dihedral angles of triangles}\label{sec2-3-1}
This section introduces a class of bending-energy discretization methods that utilize the dihedral angles between adjacent triangles to generate a curvature-free representation of the bending energy. The KN1 model, which is derived directly from a continuous formulation, serves as the foundation. The KN2 and KN3 models are subsequently introduced using simplified assumptions.
\subsubsection{KN1 model}
In the discretized form of Eq. (\ref{Eq.9}), the relationship holds using the normal vector $\bm{n}$.
\begin{equation}2\int{}^{}\Big(H^2-\frac{G}{2}\Big)dA=\frac{1}{2}\int{}^{}(\partial^{\alpha}\bm{n})\cdot(\partial_{\alpha}\bm{n})dA
\end{equation}
Therefore, the KN1 model can be used to represent the discrete form of the bending energy based on the dihedral angles between adjacent triangles as follows \cite{Gompper1996}: 
\begin{equation}\label{Eq.10}
    E_{KN1}=\tilde{K_B}\sum_{<i,j>}^{}\frac{l_{ij}}{\sqrt{3}\sigma_{ij}}[1-\cos(\theta^D_{ij}-\theta_0)]
\end{equation}
where $l_{ij}$ is the length of edge $ij$ between cells $i$ and $j$, $\sigma_{ij}={2(S_{ijk}+S_{ijl})}/({3l_{ij}})$ is the sum of one-third of the height of two triangles adjacent to edge $ij$ with $l_{ij}$ as the base \cite{Grinspun2003,Leembruggen2022}, $\theta^D_{ij}$ denotes the dihedral angle between the two adjacent triangles, and $\theta_0$ represents the equilibrium angle (Fig. \ref{Fig.2-3}(a)).

\subsubsection{KN2 model}
Assuming that all the triangles are equilateral, i.e., $l_{ij}=\sqrt{3}\sigma_{ij}$, we obtain the following equation for the KN2 model:
\begin{equation}\label{Eq.11}
    E_{KN2}=\tilde{K_B}\sum_{<i,j>}^{}[1-\cos(\theta^D_{ij}-\theta_0)]
\end{equation}

\subsubsection{KN3 model}
The KN3 model represents $\cos(\theta^D_{ij}-\theta_0)$ using terms up to the second-order of the Taylor expansion. 
\begin{equation}\label{Eq.12}
    E_{KN3}=\frac{\tilde{K_B}}{2}\sum_{<i,j>}^{}(\theta^D_{ij}-\theta_0)^2
\end{equation}

Most studies on vesicles and RBCs have used the KN2 model \cite{Tsubota2014,Pivkin2008,Zhao2012,Fedosov2014,Lykov2015}, whereas some \cite{Grinspun2003,Leembruggen2022} have used the KN1 model.

The relationship between $K_B$ and $\tilde{K_B}$ generally depends on the surface geometry of the entire triangular mesh. Additionally, $\tilde{K_B}=\sqrt{3}K_B$ for spheres \cite{Gompper1996,Kroll1992} and $\tilde{K_B}={2K_B}/{\sqrt{3}}$ for cylinders \cite{Gompper1996,Seung1988,ZZhang1995}.

However, when bending is dominant and the surface can be approximated as an equilateral triangle, $K_G = -{4K_B}/{3}$ holds for arbitrary shapes using $\tilde{K_B}={2K_B}/{\sqrt{3}}$ \cite{Schmidt2012}. As this relationship was also used in a previous study for comparing bending energies \cite{Koumoutsakos2020}, we also adopted it in this study.

\subsection{Methods using discretized curvature}\label{sec2-3-2}
We used $E=\sum_{i}^{}(H_i-H_0)^2A_i$, where $H_i$ and $A_i$ are the mean curvature and area of cell $i$, respectively. Based on the definitions of $H_i$ and $A_i$, the following ten methods elucidated in the following sections were employed. In addition to the five methods used to compare models in previous studies \cite{Guckenberger2016,Koumoutsakos2020}, we included five models originally proposed as methods for estimating the curvature of a triangular mesh \cite{Surazhsky2003,A-Review2013}.

\subsubsection{J\"{u}licher model}
\begin{figure}[p]
    \includegraphics[keepaspectratio, scale=0.5]{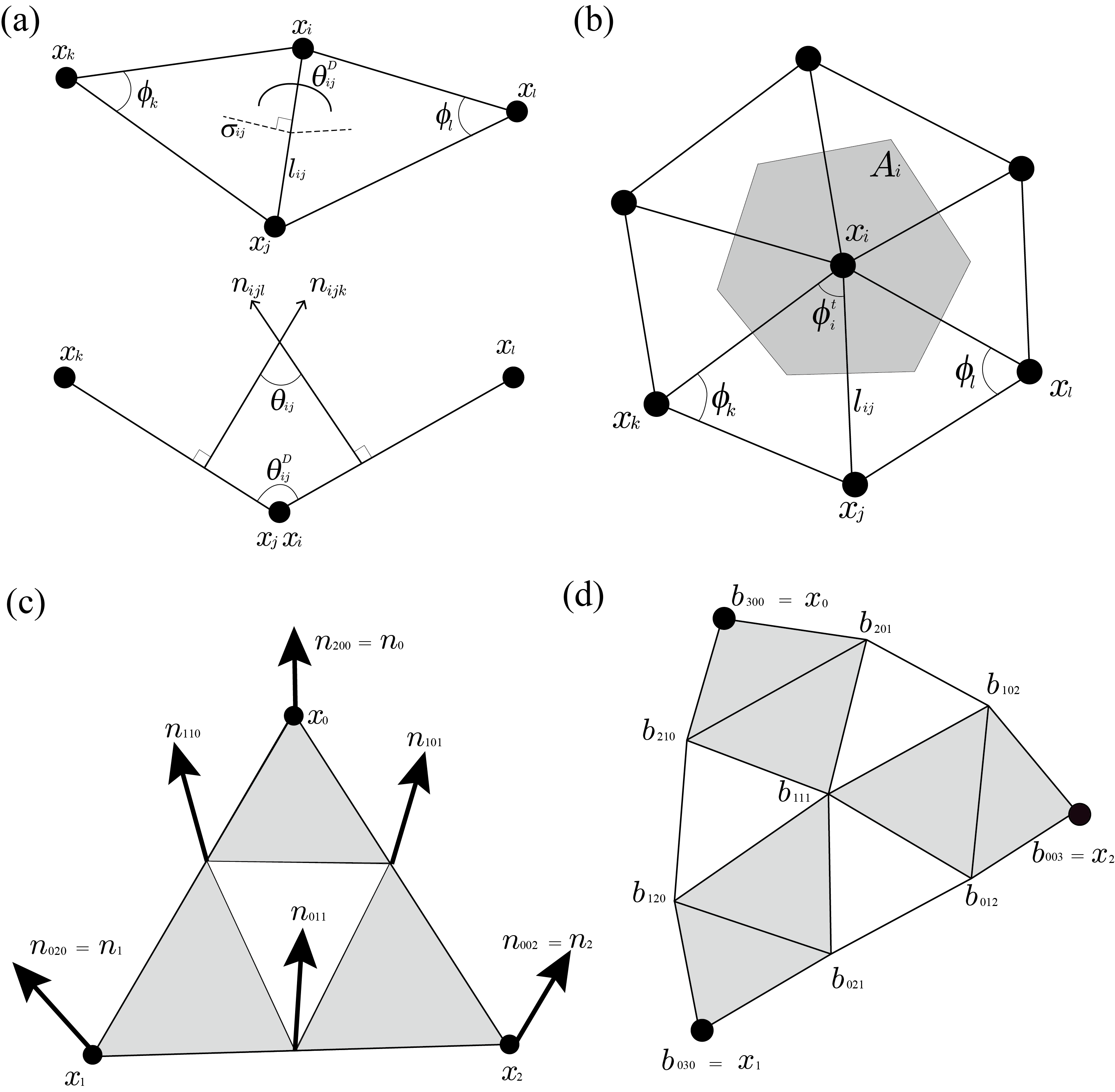}
    \caption{Variables used in the bending-energy calculations. (a) For four vertices $i$,$j$,$k$, and $l$, $l_{ij}$ is the edge between $i$ and $j$. $\sigma_{ij}$ is one-third of the sum of the heights of the two triangles with $l_{ij}$ as the base. $\phi_k$ and $\phi_l$ represent angles $ikj$ and $ilj$, respectively. $\theta^D_{ij}$ is the dihedral angle between the two triangles, and $\theta_{ij}$ is the angle between their normal vectors. (b) Cells directly connected to cell $i$ (1-ring). For a triangle around vertex $i$, $\phi^t_i$ represents the angle at vertex $i$ of the triangle adjacent to $i$. $A_i$ is the area of vertex $i$. (c) Quadratic B\'ezier surface interpolation of triangle vertex normal vectors. In the Theisel model, the mean curvature is calculated for each small gray triangle. (d) Cubic B\'ezier surface interpolation of triangle vertices. In the Vlachos model, the curvature at the gray quadrangle vertices is averaged to obtain the vertex curvature. Subsequently, it takes the weighted average of these curvatures for each adjacent triangle.}
    \label{Fig.2-3}
\end{figure}

In the J\"{u}licher model \cite{Koumoutsakos2020,Julicher1996}, the mean curvature $H^J_i$ and area $A^J_i$ are calculated using the following equations: 
\begin{equation}\label{Eq.Julicher1}
    H^J_{i}=-\frac{1}{4A^J_i}\sum_{<i,j>}^{}l_{ij}\theta_{ij},
\end{equation}
\begin{equation}\label{Eq.Julicher2}
    A^J_i=\frac{1}{3}\sum_{<i,j,k>}^{}S_{ijk},
\end{equation}
where $\theta_{ij}$ is the angle between the normal vectors, $\bm{n_{ijk}}$ and $\bm{n_{ijl}}$, of triangles $ijk$ and $ijl$, respectively, adjacent to the edge $ij$ (Fig. \ref{Fig.2-3}(a)) and $S_{ijk}$ is the area of the triangle (Fig. \ref{Fig.2-3}(b)).

\subsubsection{Watanabe model}
The mean curvature $H$ of a surface is expressed as follows:
\begin{equation}
    H =\frac{1}{2\pi} \int_{0}^{2\pi} \kappa_n(\alpha )\,d\alpha
\end{equation} 
where $\kappa_n(\alpha)$ is the normal curvature in the direction defined by angle $\alpha$ from a reference direction on the tangent plane.
The Watanabe model \cite{Surazhsky2003,Watanabe2001} approximates this relationship for a discrete surface to obtain the following equation:

\begin{equation}\label{Eq.Watanabe1}
    H^W_{i}=\frac{1}{2\pi}\sum_{<i,j>}^{}\kappa^{ij}_n\Big(\frac{\alpha_j+\alpha_{j+1}}{2}\Big)
\end{equation}
where $\kappa^{ij}_n$ is the normal curvature at vertex $i$ toward vertex $j$ and is expressed as:
\begin{equation}\label{Eq.Watanabe2}
    \kappa^{ij}_n = \frac{2\bm{n_i}\cdot (\bm{x_j}-\bm{x_i})}{\|\bm{x_j}-\bm{x_i}\|^2}
\end{equation}
where $\bm{x_i}$ and $\bm{n_i}$ are the coordinates and normal vectors at vertex $i$. The angle $\alpha_j$ represents the interior angle at vertex $i$ formed by vertices $i$, $j$, and $j+1$, after the triangle is projected onto a plane perpendicular to $\bm{n_i}$. The definition of $A_i$ is the same as that in the J\"{u}licher model, $A^J_i$ (Eq. \ref{Eq.Julicher2}).

\subsubsection{Laplace--Beltrami operator based models}
For the next three methods, the discretized Laplace--Beltrami operator was used to compute the mean curvature using the following equation:
\begin{equation}\label{Eq.12.5}
    H_i =\frac{1}{2}(\nabla^2_s \bm{x_i})\cdot \bm{n_i}
\end{equation}
where $\nabla^2_s$ denotes the Laplace--Beltrami operator.

\paragraph{Gompper--Kroll model}
\quad\\
For the GK model \cite{Gompper1996}, let $k$ and $l$ be the vertices opposite edge $ij$ in the two triangles sharing the edge. Then, $\phi_k$ and $\phi_l$ represent the interior angles $\angle ikj$ and $\angle ilj$, respectively (Fig. \ref{Fig.2-3}(b)). The discrete Laplace--Beltrami operators on the vertex positions $\nabla^2_s \bm{x_i}$ and area $A^{GK}_i$ associated with each vertex are calculated as follows: 
\begin{equation}\label{Eq.GK1}
    \nabla^2_s \bm{x_i}=-\frac{\sum_{<i,j>}^{}(\cot\phi_k+\cot\phi_l)(\bm{x_i}-\bm{x_j})}{2A^{GK}_i},
\end{equation}
\begin{equation}\label{Eq.GK2}
    A^{GK}_i=\frac{1}{8}\sum_{<i,j>}^{}(\cot\phi_k+\cot\phi_l)\|\bm{x_i}-\bm{x_j}\|^2
\end{equation}
where $A^{GK}_i$ is the sum of the areas of the quadrangles with four vertices in each triangle $ijk$: vertex $i$, midpoint of edge $ij$, circumcenter of triangle $ijk$, and midpoint of edge $ik$ for all adjacent triangles.

\paragraph{Meyer model}
\quad\\
Because Eq. (\ref{Eq.GK2}) includes a cotangent, it may adopt extreme values for obtuse triangles. Therefore, the Meyer model used $A^M_i$ instead of $A^{GK}_i$ \cite{Meyer2003}. Here, $A^M_i$ is the area calculated using the midpoint of the longest edge of triangle $ijk$ when it has an obtuse angle, instead of using the circumcenter as in $A^{GK}_i$.
In this method, the force density applied to vertex $i$ is obtained as \cite{Koumoutsakos2020,Zhong1989}:
\begin{equation}\label{Eq.force_def}
    \bm{f_i}=-2K_B[2(H_i-H_0)(H_i^2+H_0H_i-G_i)+\nabla^2_s H_i]\bm{n_i}    
\end{equation}

The force acting on vertex $i$, denoted as $\bm{F^B_i}$, is then calculated by multiplying the force density by the area:
\begin{equation}\label{Eq.force2}
    \bm{F^B_i}=\bm{f_i}A^M_i
\end{equation}
$\bm{f_i}$ is calculated as follows:
\begin{equation}\label{Eq.Meyer1}
    \nabla^2_s \bm{x_i}=-\frac{\sum_{<i,j>}^{}(\cot\phi_k+\cot\phi_l)(\bm{x_i}-\bm{x_j})}{2A^{M}_i},
\end{equation}
\begin{equation}\label{Eq.Meyer2}
    \nabla^2_s H_{i}=-\frac{\sum_{<i,j>}^{}(\cot\phi_k+\cot\phi_l)(H_i-H_j)}{2A^{M}_i},
\end{equation}
\begin{equation}\label{Eq.Meyer3}
    G^{M}_i=\frac{1}{A^{M}_i}\bigg(2\pi-\sum_{<i,j,k>}^{}\phi^t_i\bigg).
\end{equation}
where $G^M_i$ and $\phi^t_i$ denote the Gaussian curvature and interior angle at vertex $i$ of triangle $ijk$ (Fig. \ref{Fig.2-3}(b)).

\paragraph{Belkin model}
\quad\\
Similar to the Meyer model, the Belkin model \cite{Belkin2008} uses a discrete Laplace--Beltrami operator. However, it employs a different definition of this operator for the vertex positions and mean curvatures, as elucidated in the following equations:
\begin{equation}\label{Eq.Belkin1}
    \nabla^2_s \bm{x_i}=-\frac{1}{4\pi h^2}\sum_{t=1}^{N_t}\frac{S_t}{3}\sum_{\bm{p}\in V(t)}^{} \exp\Big({-\frac{\|\bm{x_i}-\bm{p}\|^2}{4h}}\Big)(\bm{x_i}-\bm{p})
\end{equation}
\begin{equation}\label{Eq.Belkin2}
    \nabla^2_s H_i=-\frac{1}{4\pi h^2}\sum_{t=1}^{N_t}\frac{S_t}{3}\sum_{\bm{p}\in V(t)}^{} \exp\Big({-\frac{\|\bm{x_i}-\bm{p}\|^2}{4h}}\Big)(H_i-H_p)
\end{equation}
where $\sum_{\bm{p} \in V(t)}$ denotes the summation over all vertices $\bm{p}$ in set $V(t)$ comprising the vertex coordinates of triangle $t$, $N_t$ is the number of triangles, and $S_t$ is the area of triangle $t$. Based on the approach employed in a previous study \cite{Guckenberger2016}, we set parameter $h$ to $A^{M}_i$. Additionally, we used Eqs. (\ref{Eq.Meyer3}) and (\ref{Eq.force2}) to calculate $G_i$ and the force, respectively.

\subsubsection{Least-squares-fitting-based models}
This section introduces three least-squares fitting-based models to define the local coordinate system at each vertex of a given surface. This approach allows us to fit the surface locally and subsequently calculate both the mean and Gaussian curvatures using differential geometry techniques. Several methods exist for local surface fitting using the least-squares technique. Although the Farutin model\cite{FARUTIN2014} employs least-squares fitting for partial derivatives with respect to the surface coordinates, alternative approaches directly fit the fundamental forms of the surface \cite{Hamann1993,Goldfeather2004}. Based on these studies, we introduced the Hamann and Goldfeather models.

\paragraph{Farutin model}
\quad\\
In Farutin model \cite{FARUTIN2014}, using unit vectors $\bm{\xi^i}$ and $\bm{\eta^i}$ perpendicular to the normal vector $\bm{n_i}$ at vertex $i$, the local coordinates of a neighboring vertex $j$ with respect to vertex $i$ are obtained using $s^j_\xi=(\bm{x_j}-\bm{x_i})\cdot \bm{\xi^i}$ and $s^j_\eta=(\bm{x_j}-\bm{x_i})\cdot \bm{\eta^i}$.

Subsequently, $\partial_\xi \bm{x_i}$, $\partial_\eta \bm{x_i}$, $\partial_{\xi\xi} \bm{x_i}$,$\partial_{\xi\eta} \bm{x_i}$, and $\partial_{\eta\eta} \bm{x_i}$ are estimated by applying the least-squares method to function $\chi$ as follows:
\begin{equation}\label{Eq.Farutin}
    \chi=\sum_{<i,j>}^{}\bigg\lvert\bigg\lvert\bm{x_j}-\bm{x_i}-\partial_\xi \bm{x_i} s^j_\xi-\partial_\eta \bm{x_i} s^j_\eta-\frac{1}{2}\Big[\partial_{\xi\xi} \bm{x_i} (s^j_\xi)^2+ \partial_{\eta\eta} \bm{x_i} (s^j_\eta)^2+2\partial_{\xi\eta} \bm{x_i} (s^j_\xi s^j_\eta)\Big]\bigg\lvert\bigg\lvert^2
\end{equation}

The normal vector is then updated using $\bm{n_i}={\partial_\xi \bm{x_i}\times \partial_\eta \bm{x_i}}/{\|\partial_\xi \bm{x_i}\times \partial_\eta \bm{x_i}\|}$. To maintain simplicity, the unit vectors $\bm{\xi^i}$ and $\bm{\eta^i}$ are not updated. The metric tensor $g^i_{\alpha\beta} = \partial_\alpha \bm{x_i}\cdot \partial_\beta \bm{x_i}$ and curvature tensor $c^i_{\alpha \beta}=\bm{n_i}\cdot \partial_{\alpha\beta}\bm{x_i}$ are computed, where $\alpha$ and $\beta$ are indices that adopt values from set $\{\xi, \eta\}$. 
The mean and Gaussian curvatures at vertex $i$ can then be calculated as follows:
\begin{equation}
H_i=\frac{1}{2}{\rm Tr} [{c^i (g^i)^{-1}}{\rm}],\quad G_i={\rm det} [c^i (g^i)^{-1}].
\end{equation}

Similarly, the discrete Laplace--Beltrami operator for the mean curvature $\nabla^2_s H$ can be expressed as:
\begin{equation}
    \nabla^2_s H=\frac{1}{\sqrt{|{\rm det} g |}}\partial_\alpha (\sqrt{|{\rm det} g|}(g^{-1})_{\alpha\beta}\partial_\beta H)
\end{equation} 
where ${\rm det} g$ denotes the determinant of the metric tensor.

$\partial_\xi H_i$, $\partial_\eta H_i$, $\partial_{\xi\xi} H_i$,$\partial_{\xi\eta} H_i$, and $\partial_{\eta\eta} H_i$ can be calculated using the least-squares method for the mean curvature and subsequently, the Laplacian can be computed as follows \cite{FARUTIN2014}:
\begin{equation}\label{Eq.Farutin2}
    \nabla^2_s H_i=\partial_{\alpha\beta}H_i(g^i)^{-1}_{\alpha\beta}-[(g^i)^{-1}_{\alpha\beta}\partial_{\alpha\beta}\bm{x_i}]\cdot[(g^i)^{-1}_{\gamma\delta}\partial_\gamma H_i \partial_\delta \bm{x_i}], \alpha,\beta,\gamma,\delta\in\{\xi,\eta\} 
\end{equation}
To calculate the force, we used Eq. (\ref{Eq.force2}).

In the least-squares method, degeneracy occurs when the number of neighboring vertices is less than five. In a previous study comparing bending models \cite{Guckenberger2016}, instances of degeneracy were handled by fitting not only the 1-ring, which considers only the first nearest neighbor vertices, but also the 2-ring vertices. However, in general, increasing the ring size requires fitting higher-order functions to represent complex surfaces, thereby significantly increasing the computational cost \cite{Ray2012,Chen2015}. Furthermore, highly curved surfaces are more likely to become multivalued functions with respect to the local coordinates, making them unsuitable for deformation simulations. Therefore, in this study, we used only the 1-ring and set $\partial_\xi \bm{x_i}=\bm{\xi^i}$ and $\partial_\eta \bm{x_i}=\bm{\eta^i}$ in cases wherein the number of neighboring vertices was less than five, and calculated the remaining three variables $\partial_{\xi\xi} \bm{x_i}$, $\partial_{\xi\eta} \bm{x_i}$, and $\partial_{\eta\eta} \bm{x_i}$ using Eq. (\ref{Eq.Farutin}). 

To prevent degeneracy in the least-squares fitting of the mean curvature, we implemented the following procedure: First, calculate the normal vector $\bm{n^h}$ using Eq. (\ref{Eq.NormalVector}) in the coordinate system $\bm{r_j}=(s^j_\xi,s^j_\eta,H_j)^T$. Given the relationships $\partial_\xi \bm{r_j} = (1,0,\partial_\xi H_j)^T$ and $\partial_\eta \bm{r_j} = (0,1,\partial_\eta H_j)^T$, $\partial_\xi \bm{r_j}\times \partial_\eta \bm{r_j}=(-\partial_\xi H_j,-\partial_\eta H_j, 1)^T$. Assuming $\bm{n^h}$ aligns in this direction, we incorporate two additional equations, $-\partial_\xi H_j {n^h}_z = {n^h}_x$ and $-\partial_\eta H_j {n^h}_z = {n^h}_y$, into the least-squares fitting procedure:

\paragraph{Hamann model}
\quad\\
In the Hamann model \cite{Hamann1993}, considering a coordinate system defined as $(s_\xi,$ $s_\eta,$ $f(s_\xi,s_\eta))$, surface fitting is performed using the following equation:
\begin{equation}
    f(s_\xi,s_\eta) = \frac{1}{2}(c_{20}s^2_\xi + 2c_{11}s_\xi s_\eta + c_{02}s^2_\eta).
\end{equation}
The coefficients $c_{20}$, $c_{11}$, and $c_{02}$ are fitting parameters determined by the surface fitting process.
In this coordinate system, $s_\xi$ and $s_\eta$ represent the local coordinates within the tangent plane, and $f(s_\xi,s_\eta)$ defines the surface height relative to this plane.

Because three variables must be determined, degeneracy occurs only at the mesh boundaries. The mean curvature $H_i$ and Gaussian curvature $G_i$ are expressed as follows:
\begin{equation}
    H_i = c_{20}c_{02}-c_{11}^2,\quad G_i = \frac{1}{2}(c_{20}+c_{02}).
\end{equation}
In the local coordinate system $(s_\xi,s_\eta,f(s_\xi,s_\eta))^T$, the metric tensor at vertex $i$ becomes the identity tensor. Additionally, at vertex $i$, the first-order partial derivative terms with respect to the local coordinates have a zero $z$ component, whereas the second-order partial derivative terms have zero $x$ and $y$ components. Consequently, the inner products of Eq. (\ref{Eq.Farutin2}) becomes zero and it can be simplified as
\begin{equation}
    \nabla^2_s H_i=\partial_{\xi\xi}H_i+\partial_{\eta\eta}H_i
\end{equation}
The fitting of the partial derivative terms of the mean curvature follows the same procedure as that of the Farutin model. Finally, the force calculations are performed using Eq. (\ref{Eq.force2}).

\paragraph{Goldfeather model}
\quad\\
In the Goldfeather model \cite{Goldfeather2004}, surface fitting is performed using the following equation: 
\begin{multline}
    f(s_\xi,s_\eta) = c_{10} s_\xi + c_{01} s_\eta + \frac{1}{2}\Big(c_{20}s^2_\xi + 2c_{11}s_\xi s_\eta + c_{02}s^2_\eta\Big) \\
    + c_{30}s^3_\xi + c_{21}s^2_\xi s_\eta + c_{12}s_\xi s^2_\eta + c_{03}s^3_\eta
\end{multline}
The coefficients $c_{ij}$ are fitting parameters determined by the surface fitting process.
Using only the 1-ring neighborhood, two additional equations are incorporated for each vertex based on the normal vector information $n(s_\xi,s_\eta)$ and $f(s_\xi,s_\eta)$:
\begin{equation}\label{Eq.Goldfeather}
\bm{n}(s_\xi,s_\eta) = -\begin{pmatrix}
    c_{10}+c_{20}s_\xi+c_{11}s_\eta+3c_{30}s^2_\xi+2c_{21}s_\xi s_\eta+c_{12}s^2_{\eta}\\
    c_{01} +c_{11}s_\xi+c_{02}s_\eta+c_{21}s^2_\xi+2c_{12}s_\xi s_\eta+3c_{03}s^2_\eta\\
    -1
    \end{pmatrix}.
\end{equation}

By imposing the condition that the normal vector $\bm{n_j}=(a_j,b_j,c_j)^T$ obtained from Eq. (\ref{Eq.NormalVector}) is parallel to $\bm{n}(s_\xi,s_\eta)$, we can formulate $3N_i$ equations for $N_i$ vertices adjacent to vertex $i$. Therefore, degeneracy occurs only at the mesh boundaries.

The mean curvature $H_i$ and Gaussian curvature $G_i$ can be calculated using $H_i={\rm Tr} [{c^i (g^i)^{-1}}{\rm}]/2$ and $G_i={\rm det} [c^i (g^i)^{-1}]$, respectively. Matrices $g$ and $c$ are calculated as follows:

\begin{equation}
g = \begin{pmatrix}
    1+c_{10}^2 & c_{10}c_{01} \\
    c_{10}c_{01} & 1+c_{01}^2 \\
    \end{pmatrix},\quad
c = \frac{1}{\sqrt{1+c_{10}^2+c_{01}^2}}
    \begin{pmatrix}
    c_{20} & c_{11}\\
    c_{11} & c_{02} \\
    \end{pmatrix}.
\end{equation}

Subsequently, the normal vector is updated using Eq. (\ref{Eq.Goldfeather}), and the local coordinate system is redefined such that $g$ becomes an identity tensor. The discrete Laplace--Beltrami operator on the mean curvatures $\nabla^2_s H_{i}$ is then calculated in the same manner as in the Hamann model. Finally, the force calculations are performed using Eq. (\ref{Eq.force2}).

\subsubsection{Triangle-based models}
Next, we describe two triangle-based models: Theisel \cite{Theisel2004} and Vlachos \cite{Vlachos2001}. Both models conduct interpolation within a triangle using the coordinates of the vertices $\bm{x_0}$, $\bm{x_1}$, and $\bm{x_2}$ and the corresponding unit normal vectors $\bm{n_0}$, $\bm{n_1}$, and $\bm{n_2}$.

\paragraph{Theisel model}
\quad\\
The Theisel model estimates the mean curvature $H(u, v)$ and Gaussian curvature $G(u, v)$ at a point $(u, v)$ within the triangle. Here, $(u, v)$ represents a local barycentric coordinate system within a triangle, where $(0,0)$, $(1,0)$, and $(0,1)$ correspond to vertices $\bm{x_0}$, $\bm{x_1}$, and $\bm{x_2}$, respectively.

$H(u, v)$ and $G(u, v)$ are calculated as follows:
\begin{equation}
    H(u,v) = \frac{1}{2} \frac{\bm{\tilde{n}}\cdot \bm{h}}{\|\bm{\tilde{n}}\|(\bm{\tilde{n}}\cdot \bm{\tilde{m}})},
\end{equation}
\begin{equation}
    G(u,v) = \frac{{\rm det} (\bm{n_0},\bm{n_1},\bm{n_2})}{\|\bm{\tilde{n}}\|^2(\bm{\tilde{n}}\cdot \bm{\tilde{m}})},
\end{equation}
where
\begin{subequations}
    \begin{equation}
        \bm{\tilde{n}} = (1-u-v)\bm{n_0} +u \bm{n_1} + v\bm{n_2},
    \end{equation}
    \begin{equation}
        \bm{\tilde{m}} = \bm{r_2}\times \bm{r_0},
    \end{equation}
    \begin{equation}
        \bm{h} = (\bm{n_0}\times \bm{r_0})+(\bm{n_1}\times \bm{r_1})+ (\bm{n_2}\times \bm{r_2}),
    \end{equation}
    \begin{equation}
        \bm{r_0}=\bm{x_2}-\bm{x_1}, \quad  \bm{r_1}=\bm{x_0}-\bm{x_2}, \quad \bm{r_2} = \bm{x_1}-\bm{x_0}.
    \end{equation}
\end{subequations}
${\rm det} (\bm{n_0},\bm{n_1},\bm{n_2})$ represents the determinant of the matrix formed by the vertex normal vectors.

Theisel et al.\cite{Theisel2004} applied this algorithm to a triangular mesh unit; however, its reliance on linear interpolation renders it ineffective for high-wavenumber. Therefore, we used it to interpolate normal vectors with the B\'ezier curve proposed by Vlachos et al.\cite{Vlachos2001} and performed the calculation for each smaller triangle within the original triangle (Fig. \ref{Fig.2-3}(c)).
The normal vector is expressed as follows:
\begin{equation}\label{Eq.bezierNormal}
    \bm{n}(u,v) = \sum_{i+j+k=2} \bm{n_{ijk}}\frac{2!}{i!j!k!}u^iv^j (1-u-v)^k.
\end{equation}
where $\sum_{i+j+k=2}$ denotes the sum over all possible combinations of nonnegative integers $i$, $j$, and $k$ that sum to 2.

The coefficients $\bm{n_{ijk}}$, which are referred to as the control points in Eq. (\ref{Eq.bezierNormal}) are calculated as follows:
\begin{subequations}
    \begin{equation}
        \bm{n_{200}} = \bm{n_0},
    \end{equation}
    \begin{equation}
        \bm{n_{020}} = \bm{n_1},
    \end{equation}
    \begin{equation}
        \bm{n_{002}} = \bm{n_2},
    \end{equation}
    \begin{equation}
        \bm{n_{110}} = \frac{\bm{h_{110}}}{\|\bm{h_{110}}\|},
    \end{equation}
    \begin{equation}
        \bm{n_{011}} = \frac{\bm{h_{011}}}{\|\bm{h_{011}}\|},
    \end{equation}
    \begin{equation}
        \bm{n_{101}} = \frac{\bm{h_{101}}}{\|\bm{h_{101}}\|},
    \end{equation}
\end{subequations}
where
\begin{subequations}
    \begin{gather}
        \bm{h_{110}} = \bm{n_0} + \bm{n_1} -v_{01}(\bm{x_1}-\bm{x_0}),\\
        \bm{h_{011}} = \bm{n_1} + \bm{n_2} -v_{12}(\bm{x_2}-\bm{x_1}),\\
        \bm{h_{101}} = \bm{n_2} + \bm{n_0} -v_{20}(\bm{x_0}-\bm{x_2}),\\
        v_{ij} = 2\frac{(\bm{x_j}-\bm{x_i})\cdot(\bm{n_i}+\bm{n_j})}{(\bm{x_j}-\bm{x_i})\cdot (\bm{x_j}-\bm{x_i})}.
    \end{gather}
\end{subequations}
Subsequently, linear interpolation is used for the triangles separated by the midpoint of each side.
For each vertex $i$, we first estimate $H(u,v)$ and $G(u,v)$ within each triangle-sharing vertex $i$. These estimates are then weighted and averaged to obtain the final estimates of $H_i$ and $G_i$ at the vertex. The weights are $A_i^M$ in the Meyer model by each triangle \cite{Rusinkiewicz2004,Zhihong2011}. The discrete Laplace--Beltrami operator on the mean curvature, $\nabla^2_s H_{i}$, is calculated using the same approach as that in the Meyer model. Finally, the force is calculated using Eq. (\ref{Eq.force2}).

\paragraph{Vlachos model}
\quad\\
The Vlachos model\cite{Vlachos2001} employes cubic B\'ezier curves for direct interpolation within triangles. A surface function $\bm{f}(u, v)$ with parameters $(u, v)$ is defined as follows:
\begin{equation}\label{Eq.surface}
    \bm{f}(u,v) = \sum_{i+j+k=3} \bm{b_{ijk}}\frac{3!}{i!j!k!}u^iv^j (1-u-v)^k,
\end{equation}
where $\bm{b_{ijk}}$ denotes the coefficient vectors that determine the surface shape, and ${i+j+k=3}$ indicates that we are summing all possible combinations of nonnegative integers $i$, $j$, and $k$ that add up to 3. The coefficient vectors $\bm{b_{ijk}}$ are calculated as follows:
\begin{subequations}
    \begin{gather}
        \bm{b_{300}} =\bm{x_0}, \quad \bm{b_{030}} =\bm{x_1},\quad \bm{b_{003}} =\bm{x_2},\\
        \bm{b_{210}} = \frac{(2\bm{x_0}+\bm{x_1}-w_{01}\bm{n_0})}{3},\\
        \bm{b_{120}} = \frac{(2\bm{x_1}+\bm{x_0}-w_{10}\bm{n_1})}{3},\\
        \bm{b_{021}} = \frac{(2\bm{x_1}+\bm{x_2}-w_{12}\bm{n_1})}{3},\\
        \bm{b_{012}} = \frac{(2\bm{x_2}+\bm{x_1}-w_{21}\bm{n_2})}{3},\\
        \bm{b_{102}} = \frac{(2\bm{x_2}+\bm{x_0}-w_{20}\bm{n_2})}{3},\\
        \bm{b_{201}} = \frac{(2\bm{x_0}+\bm{x_2}-w_{02}\bm{n_0})}{3},\\
        \bm{b_{111} = E + \frac{(E-V)}{2}},
    \end{gather}
\end{subequations}
where
\begin{subequations}
    \begin{gather}
        w_{ij} = (\bm{x_j}-\bm{x_i})\cdot \bm{n_i},\\
        \bm{E} =\frac{\bm{b_{210}}+\bm{b_{120}}+\bm{b_{021}}+\bm{b_{012}}+\bm{b_{102}}+\bm{b_{201}}}{6},\\
        \bm{V} =\frac{\bm{b_{300}}+\bm{b_{030}}+\bm{b_{003}}}{3}.
    \end{gather}
\end{subequations}
The resulting surface functions are then partially differentiated to compute the first and second fundamental forms, as well as the mean and Gaussian curvatures at the control points \cite{Zhihong2011}. Thereafter, a weighted average is taken at the vertices, similar to the Vlachos model. However, to account for potential steep changes in the B\'ezier surface, a simple average of the values at the four control points near the vertices of the triangle is computed before the weighted average (Fig. \ref{Fig.2-3}(d)). The discrete Laplace--Beltrami operator on the mean curvatures $\nabla^2_s H_{i}$ is calculated using the same approach as that in the Meyer model. Finally, the force is calculated using Eq. (\ref{Eq.force2}).

\section{Results}\label{sec3}
In the cell-center model, the vertices of the triangular mesh correspond to individual cells. Therefore, not only the overall shape of the sheet but also the arrangement of the vertices (cells) within it plays a crucial role. To comprehensively compare the bending models, considering both the shape and cell arrangement, we conducted two types of comparisons: static and dynamic. 

In Section \ref{sec3-1}, we compare the calculated energy and vertex-force density with the theoretical values using simple mesh shapes. Subsequently, in Section \ref{sec3-2}, we simulate the formation of wrinkles on a planar sheet via cell division. Because the theoretical values for energy and force are generally not obtainable in this dynamic scenario, we compared the simulation results with the theoretical relationship between the wavenumber of wrinkles on a plane and the out-of-plane constraint energy coefficient $K_Z$. Because this relationship generally holds only for wrinkles on a surface with zero curvature \cite{Stoop2015}, we selected a planar sheet as the initial configuration.

\begin{figure}[p]
    \centering
    \includegraphics[keepaspectratio, scale=0.5]
    {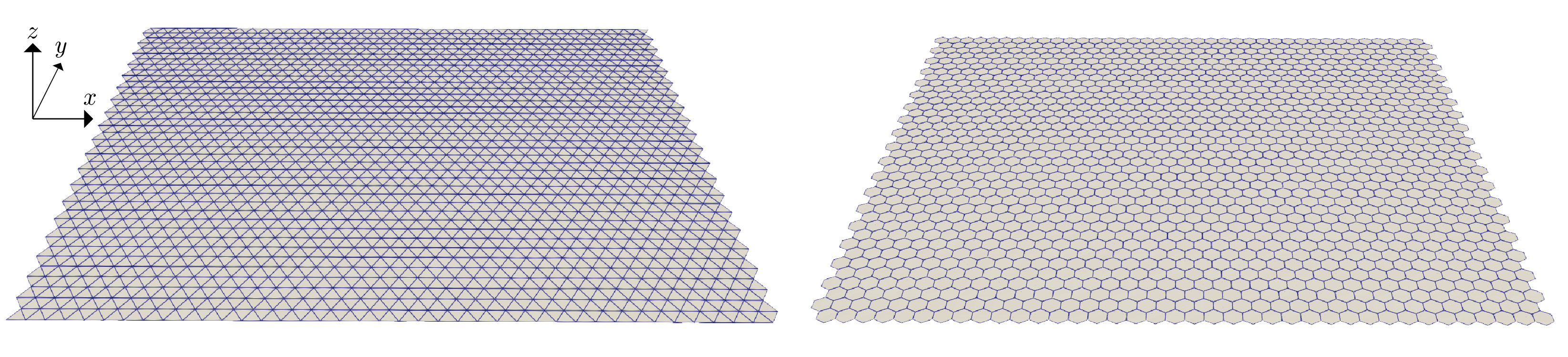}
    \caption{Initial cell-sheet configuration: it comprises 40 cells each in the $x$- and $y$-directions, for a total of 1600 cells. The left panel shows the triangular mesh and the right panel shows the 2D cell shape. The distance between each cell is $L_{\rm{eq}}$.}
    \label{Fig.3-1}
\end{figure}

Based on previous studies \cite{MIMURA2023,Inoue2020}, we prepared a planar cell sheet comprising 1600 cells ($40 \times 40$) with an intercellular distance $L_{\rm{eq}}$ on the $z=0$-plane, as illustrated in Fig. \ref{Fig.3-1}. The left panel of Fig. \ref{Fig.3-1} shows the triangular mesh representation, whereas the right panel displays the 2D cell shapes. Periodic boundary conditions were applied along both the $x$ and $y$ axes with dimensions $L_x=40L_{\rm{eq}}$ and $L_y=20\sqrt{3}L_{\rm{eq}}$, respectively.

\subsection{Static Comparison}\label{sec3-1}
We considered wrinkled planar and smooth spherical surfaces as simple curved-surface geometries for which the theoretical values of the bending energy and force density can be calculated. The errors in energy $\epsilon_E$ and force density $\epsilon_f$ are defined as follows:

\begin{equation}
    \epsilon_E = \frac{|E^a_B-E^n_B|}{E^a_B}
\end{equation}
\begin{equation}
    \epsilon_f = \frac{1}{N}\sum_{i}|f^a_i-f^n_i|
\end{equation}
where $a$ and $n$ denote the theoretical and numerical solutions, respectively, $i$ is the index of each cell, and $N$ is the total number of cells.

For $E_B$, both terms in Eq. (\ref{Eq.9}) are relevant for models using triangular dihedral angles, whereas only the first term applies to other models. $f_i$ corresponds to the magnitude of the force density $\bm{f_i}$ defined in Eq. (\ref{Eq.force_def}).

\subsubsection{Accuracy evaluation for wrinkled planar meshes}
To assess the accuracy of the bending models for a wrinkled plane, we considered the case wherein the entire sheet forms a curved surface defined by the following equations, as shown in Fig. \ref{Fig.3-2}(left).
\begin{equation}
    z(x,y)=\cos\Big[{\frac{2\pi n_x}{L_x}x}\Big]\cos\Big[{\frac{2\pi n_y}{L_y}y}\Big]
\end{equation}
Fig. \ref{Fig.3-2} shows the surfaces when (a) $n_x=4,n_y=0$, (b) $n_x=0,n_y=4$, and (c) $n_x=4,n_y=4$. Assuming that each surface is formed by cell division from an unwrinkled plane, as shown in Fig. \ref{Fig.3-1}, each cell division corresponds to (a) $x$-directional, (b) $y$-directional, and (c) random directional divisions. The degrees of freedom of the cell arrangement are then determined by the values of $M_x$ and $M_y$ in a cell sheet with $40M_x$ and $40M_y$ cells in the $x$- and $y$-directions, respectively, totaling $1600M_xM_y$ cells. In Case (a), the value of $M_x$ changes when $M_y=1$; in Case (b), the value of $M_y$ changes when $M_x=1$; and in Case (c), the value of $M_x$($M_y$) changes when $M_x=M_y$. To examine the trend in the direction of the division axis, the energy and force densities were calculated for 16 different cell configurations by varying either $M_x$ or $M_y$ from 1--4 in increments of 0.2, while keeping the other parameter constant ($M_y$ = 1 for Case (a), $M_x$ = 1 for Case (b), and $M_x = M_y$ for Case (c)).

Fig. \ref{Fig.3-2} shows the bending-energy (middle column) and force-density (right column) errors for each wrinkle pattern. Each value represents the average of the results for each of the 16 cell configurations. No significant differences can be observed in the accuracies of all models except for the dihedral, Watanabe, and Belkin models. Additionally, the KN1 model exhibits comparable accuracy to the other discretized curvature models only for the (b) $y$-directed division.

\subsubsection{Effects of mesh structure on bending model accuracy}
The previous comparisons were conducted using a structured mesh with aligned cells. To investigate the potential differences in accuracy owing to the mesh structure, we performed the same comparison using an unstructured mesh with broken symmetry. Restricting our analysis to isotropic meshes ($M_x=M_y$), we generated meshes using the equation $4\Big[\sqrt{(x^2+y^2)/{M_x^2}}-{20}/{\pi}\Big]^2+\Big(z/{M_x}\Big)^2=6.78558^2$ in pymeshlab \cite{pymeshlab}. The mesh was generated using the Isotropic Explicit Remeshing \cite{IsotorpicRemeshing} filter with a target edge length of $L_{\rm{eq}}$, followed by a coordinate transformation.

Fig. \ref{Fig.3-3} compares the energy and force-density errors of various models for both the structured and unstructured meshes. Specifically, Fig. \ref{Fig.3-3}(a) presents the results for the structured mesh shown in Fig. \ref{Fig.3-2}(c), whereas Fig. \ref{Fig.3-3}(b) presents those for the unstructured mesh. Evidently, the latter six models produced extremely small force-density errors for the unstructured mesh. By contrast, the J\"{u}licher, GK, and Meyer models produced larger force-density errors than the other discrete curvature models for the unstructured mesh but exhibited comparable accuracies for the structural mesh. Additionally, among the dihedral-angle models, the KN1 model exhibited the smallest error for the structured mesh but the largest for the unstructured mesh.

\subsubsection{Accuracy comparison for isotropic and anisotropic spherical meshes}
To further evaluate the accuracy of the bending models, we compared their performances for isotropic (Fig. \ref{Fig.3-4}(a)) and anisotropic (Fig. \ref{Fig.3-4}(b)) meshes generated on a sphere.
For the isotropic meshes, a sphere of radius $14.3M_x$ was created using pymeshlab \cite{pymeshlab} with $M_x=M_y$ and a target length of $L_{\rm{eq}}$ was used in the Isotropic Explicit Remeshing filter\cite{IsotorpicRemeshing}. Additionally, scale transformations of $1/M_x$ were applied to the $x$, $y$, and $z$ axes. For the anisotropic meshes, an ellipsoid ${x^2}/{{M_x}^2}+y^2+z^2=14.3^2$ was created using pymeshlab with $M_y=1$. Subsequently, $1/M_x$-scale transformation was applied only in the $x$ direction. 

Under these conditions, the Watanabe and Belkin models exhibited small errors, as shown in Fig. \ref{Fig.3-4}. By contrast, the J\"{u}licher and Vlachos models resulted in relatively larger errors than those for the planar mesh (Fig. \ref{Fig.3-3}). Moreover, the GK model produced a particularly large force-density error than the other models.

\subsection{Dynamic conditions}\label{sec3-2}
To compare the aforementioned bending models under dynamic conditions, we conducted simulations wherein the cell sheets (Fig. \ref{Fig.3-1}) were deformed out-of-plane via cell division. Particularly, to focus on triangular isotropy, we performed two types of simulations: one with uniaxial cell division (in the $x$ direction) and the other with random division directions. However, the Belkin model was excluded from these simulations owing to its computational cost of $O(N^2)$ for $N$ cells.

As initial conditions, small displacements were introduced by assigning $z$-coordinates to each cell using uniform random numbers in the range $-{L_{\rm{eq}}}/{100}$--${L_{\rm{eq}}}/{100}$. 

To quantify the wrinkle structure after the out-of-plane deformation, the wavenumber of the wrinkle $\overline{u}$ was used as the index and calculated using the discrete Fourier transform \cite{MIMURA2023,Inoue2020}. The out-of-plane displacement of the cell sheets was represented by $Z_{\rm{range}} = \max_i{z_i}-\min_i{z_i}$, which is the difference between the largest and smallest $z$-coordinates among all the cells. The wrinkle wavenumber was calculated when $Z_{\rm{range}}$ reached $\sqrt{2}L_{\rm{eq}}$.

The remaining parameters are listed in Table \ref{table1} based on the previous research \cite{MIMURA2023}. The Euler method was used to numerically compute the equations of motion.
Nine different values, \{$10^{-3}, 10^{-2}, 10^{-1}, 0.2, 0.4, 0.8, 1, 2, 4$\}, for the out-of-plane deformation constraint-energy coefficient, $K_Z$, were used in each simulation.

Figs. \ref{Fig.3-5}(a)--(d) show the wrinkle structures after out-of-plane deformation at $t=\tau_{i}^{\rm{cycle}}$ when cells were allowed to divide in the $x$ direction with $K_Z=0.2$.
Figs. \ref{Fig.3-5}(a')--(d') show the corresponding results for the random direction division.
Although no significant differences are evident in the wrinkle structures across the models, some variations are apparent. The KN1 model (Fig. \ref{Fig.3-5}(a)) exhbits four complete wrinkles, the J\"{u}licher model (Fig. \ref{Fig.3-5}(b)) exhibits three to four wrinkles, and the Hamann and Vlachos models (Figs. \ref{Fig.3-5}(c) and (d)) exhibit three wrinkles each. Additionally, the results of the Hamann (Fig. \ref{Fig.3-5}(c)) and Vlachos (Fig. \ref{Fig.3-5}(d)) model for the $x$-directed division, and those of the J\"{u}licher (Fig. \ref{Fig.3-5}(b')) and Vlachos (Fig. \ref{Fig.3-5}(d')) model for the random-directed division are very similar.

To quantify these observations, we investigated the relationship between the out-of-plane deformation constraint energy coefficient $K_Z$ and wavenumber of wrinkles $\overline{u}$. The log--log plots for the $x$- and random-directed divisions are shown in Figs. \ref{Fig.3-6}(a) and (b), respectively, wherein each plot point represents the average of the simulation results for five random seeds. The dotted line is a straight line with a slope that follows $\overline{u}\propto K_Z^{0.25}$, which corresponds to the theoretical relationship in a continuum \cite{Cerda2003,Brau2013}. In both models, the wavenumber is constant at $K_Z\leq10^{-2}$ and depends on $K_Z$ at $K_Z\ge10^{-1}$.

Finally, Fig. \ref{Fig.3-7} shows the relationship between the elapsed time for one calculation and the number of mesh vertices.

\section{Discussion}\label{sec4}
\subsection{Static condition}\label{sec4-1}
\begin{figure}[p]
    \centering
    \includegraphics[keepaspectratio, scale=0.7]
    {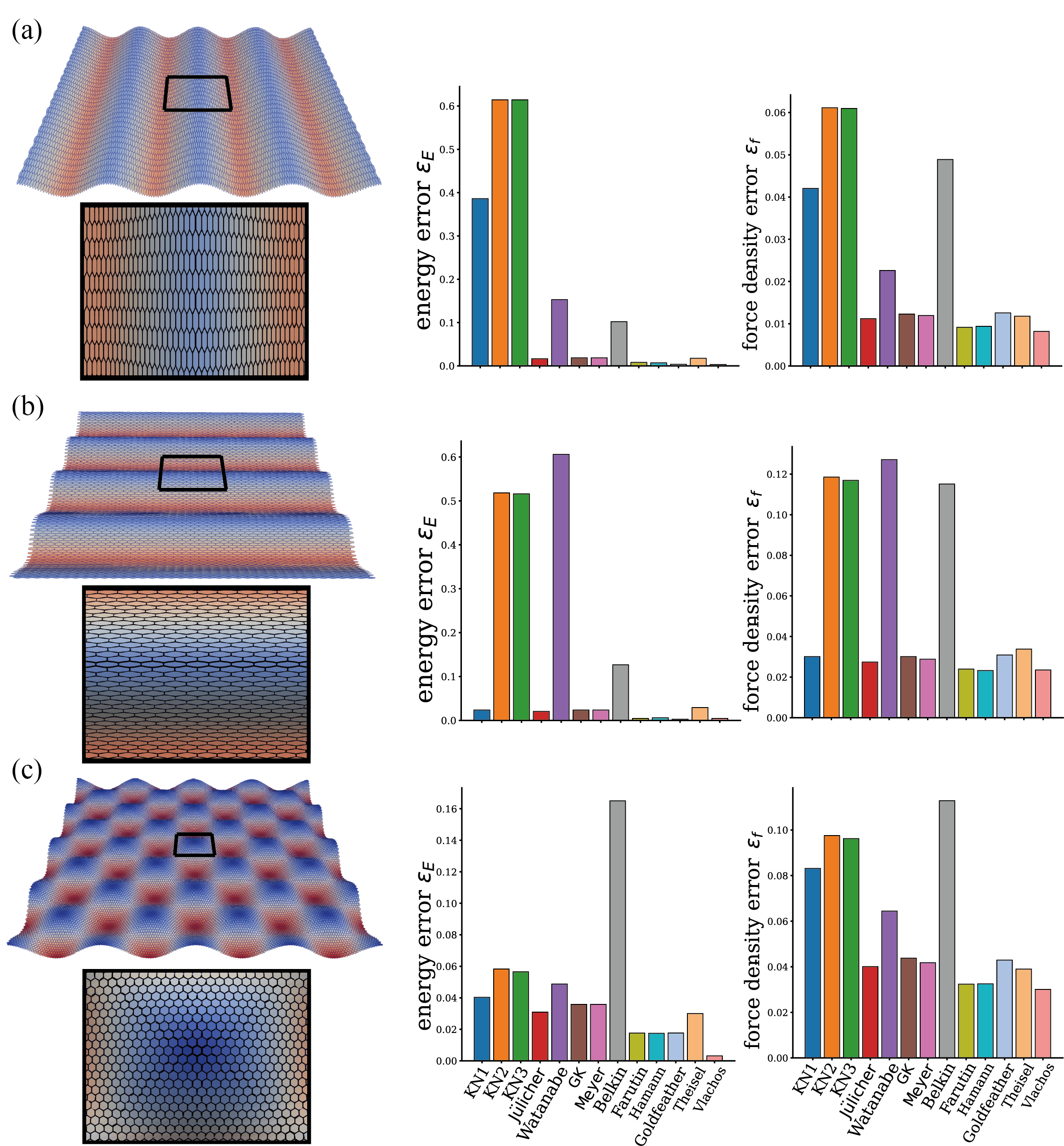}
    \caption{(left) Sheets representing surfaces following $z(x,y)=\rm{cos}\bigg[\mathit{\frac{2\pi n_x}{L_x}x}\bigg]\rm{cos}\bigg[\mathit{\frac{2\pi n_y}{L_y}y}\bigg]$. 
    The magnified view of the area enclosed by the rectangle is shown immediately below. Each magnified view represents a perspective perpendicular to the sheet.
    Surfaces for (a) $n_x=4$, $n_y=0$, (b) $n_x=0$, $n_y=4$, and (c) $n_x=4$, $n_y=4$. (right) Energy ($\epsilon_E$) and force-density ($\epsilon_f$) errors of each model.}
    \label{Fig.3-2}
\end{figure}

\begin{figure}[p]
    \centering
    \includegraphics[keepaspectratio, scale=0.8]
    {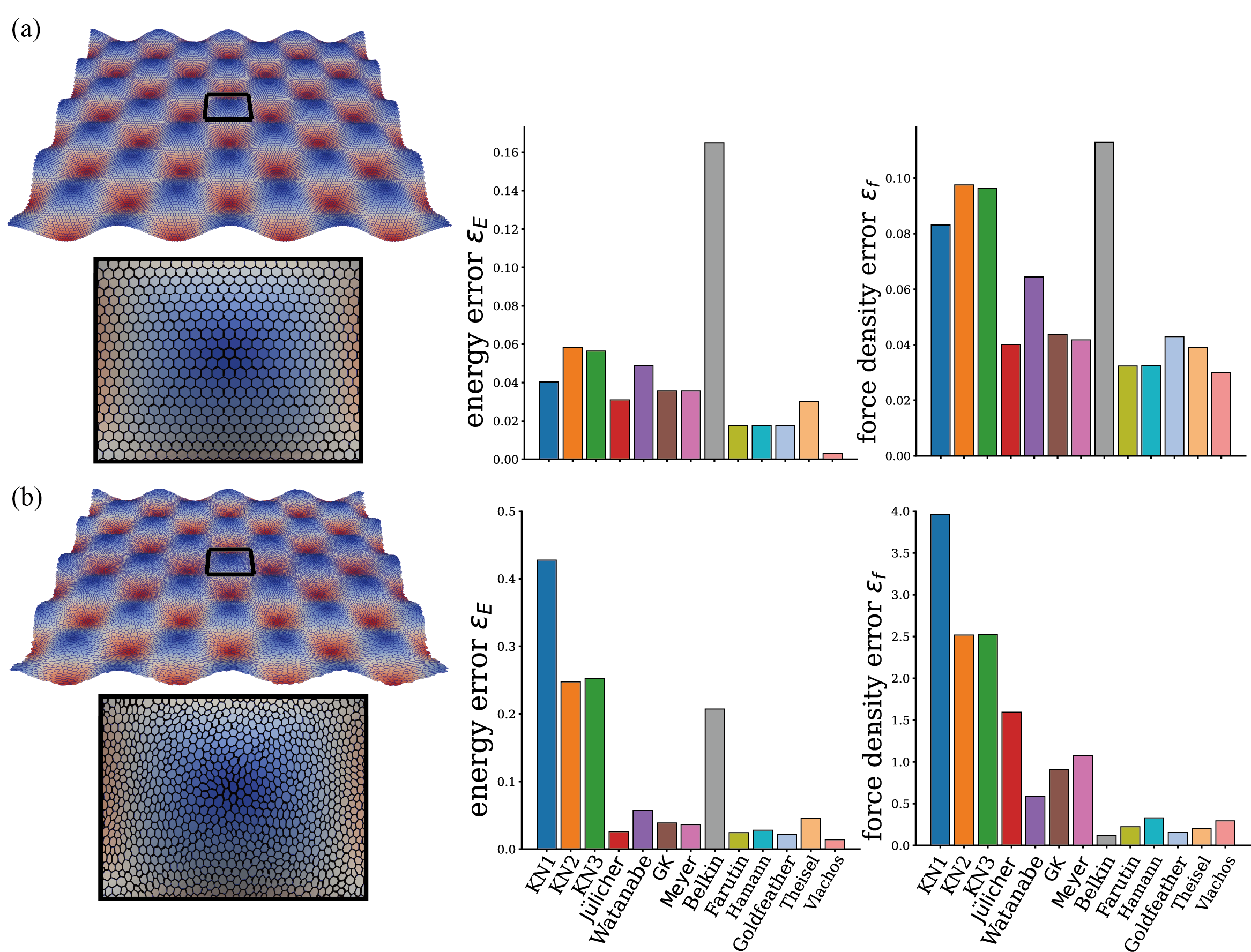}
    \caption{(a) structured and (b) unstructured meshes with $M_x=M_y=4$. The right-hand side panels illustrate the energy ($\epsilon_E$) and force-density errors ($\epsilon_f$) for each model.}
    \label{Fig.3-3}
\end{figure}

\begin{figure}[p]
    \centering
    \includegraphics[width=\textwidth, height=0.9\textheight, 
    keepaspectratio]
    {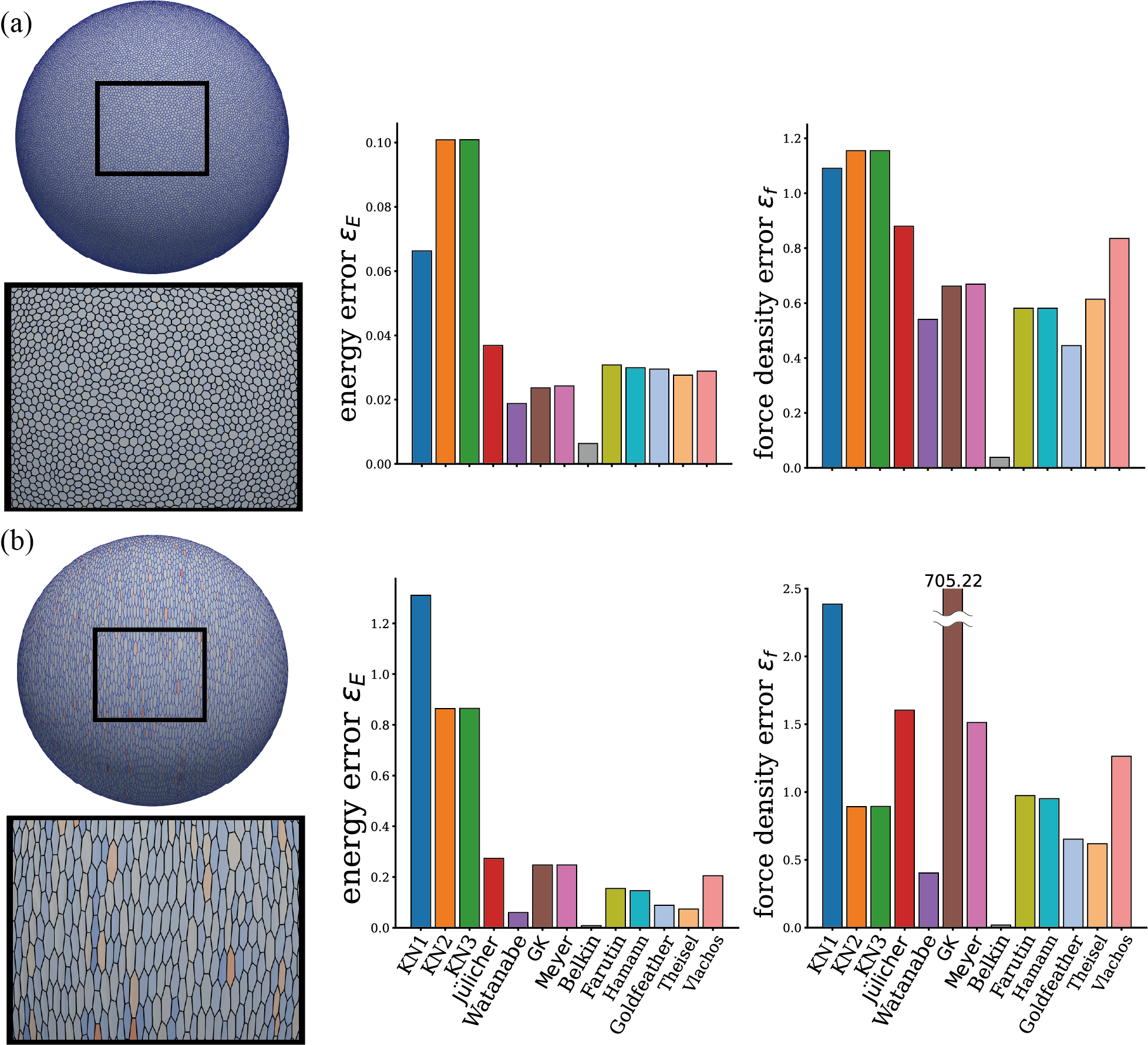}
    \caption{
        Spherical (a) isotropic and (b) anisotropic meshes with $M_x=4$. The right-hand side panels illustrate the energy ($\epsilon_E$) and force-density errors ($\epsilon_f$) for each model.
    }
    \label{Fig.3-4}
\end{figure}

\begin{figure}[p]
    \centering
    \includegraphics[width=\textwidth, height=0.9\textheight, 
    keepaspectratio]
    {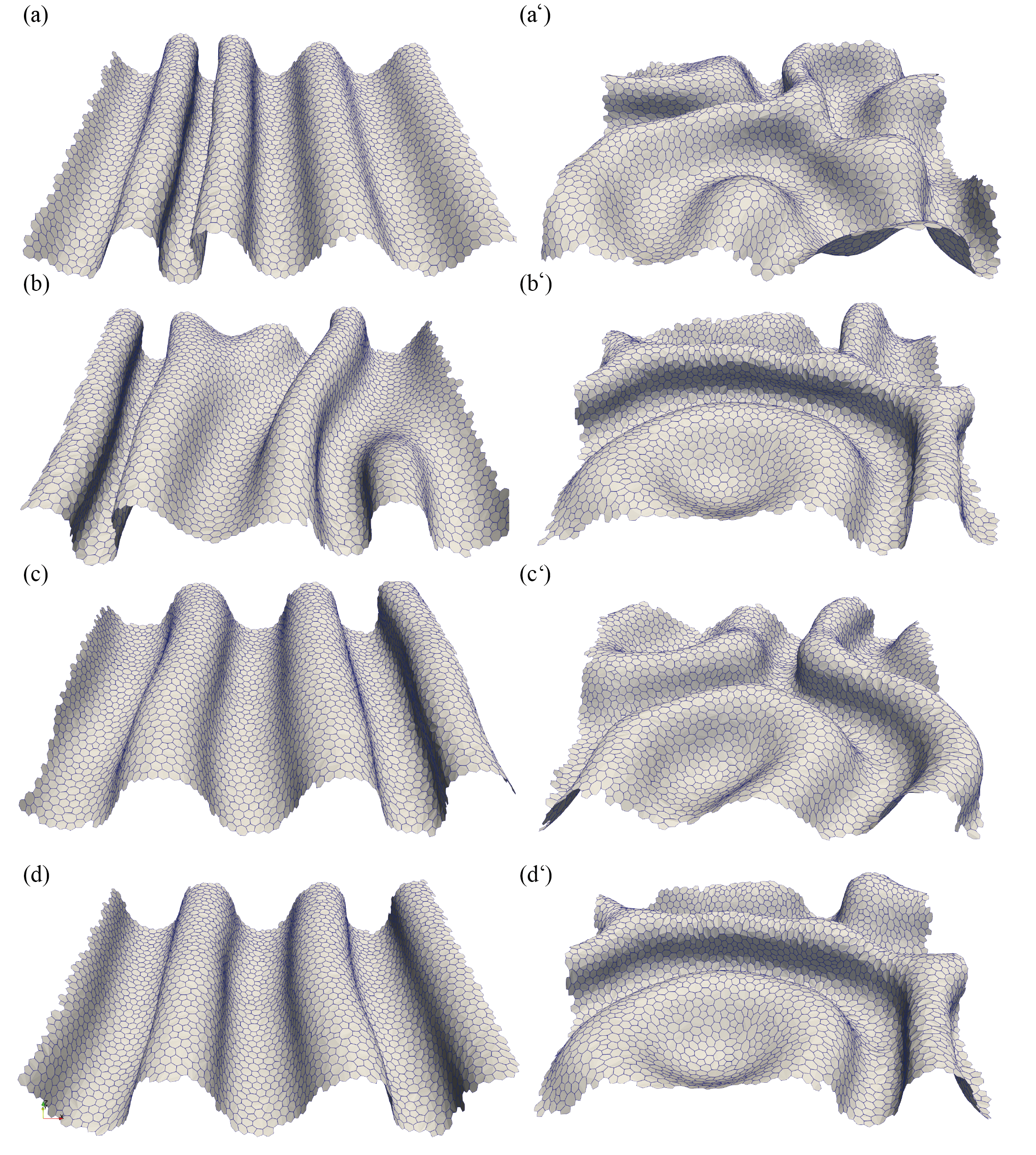}
    \caption{Snapshots taken at time $t=\tau_{i}^{\rm{cycle}}$ under an out-of-plane deformation constraint of $K_Z=0.2$. Sheets subjected to cell division (a--d) along the $x$-axis and (a'--d') random division using the (a, a') KN1, (b, b') J\"{u}licher, (c, c') Hamann, and (d, d') Vlachos models.}
    \label{Fig.3-5}
\end{figure}

\begin{figure}[p]
    \centering
    \includegraphics[width=\textwidth, height=0.9\textheight, 
    keepaspectratio]
    {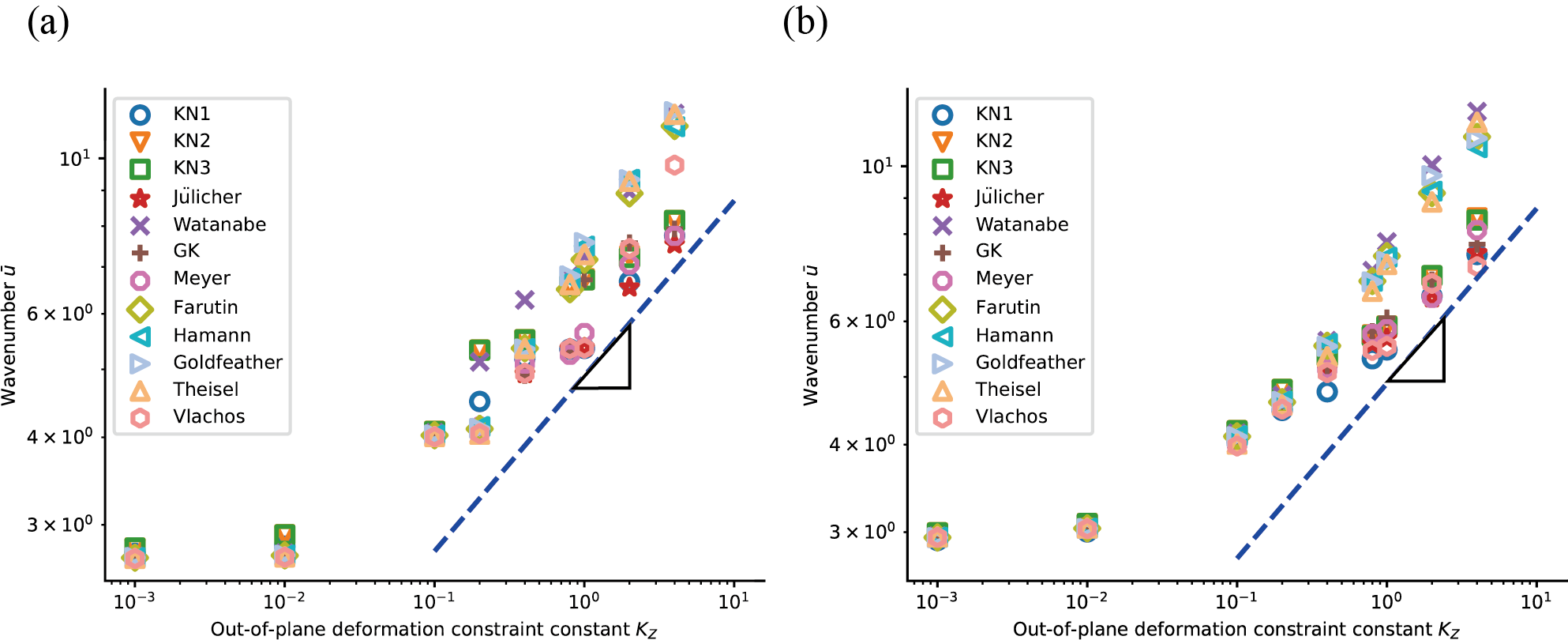}
    \caption{Relationship between the out-of-plane deformation constraint -energy coefficient $K_Z$ and the wavenumber of the folding $\overline{u}$. The dotted line represents $\overline{u}\propto K_Z^{0.25}$. (a) $x$-axis and (b) random axis division. Each data point is the average of five simulations. For $x$-axis division using the Vlachos model, data points for $K_Z = 0.8$ and $1$ are averaged over four simulations, as one simulation for each case resulted in outliers due to excessive deformation.}
    \label{Fig.3-6}
\end{figure}

\begin{figure}[p]
    \centering
    \includegraphics[width=\textwidth, height=0.9\textheight, 
    keepaspectratio]
    {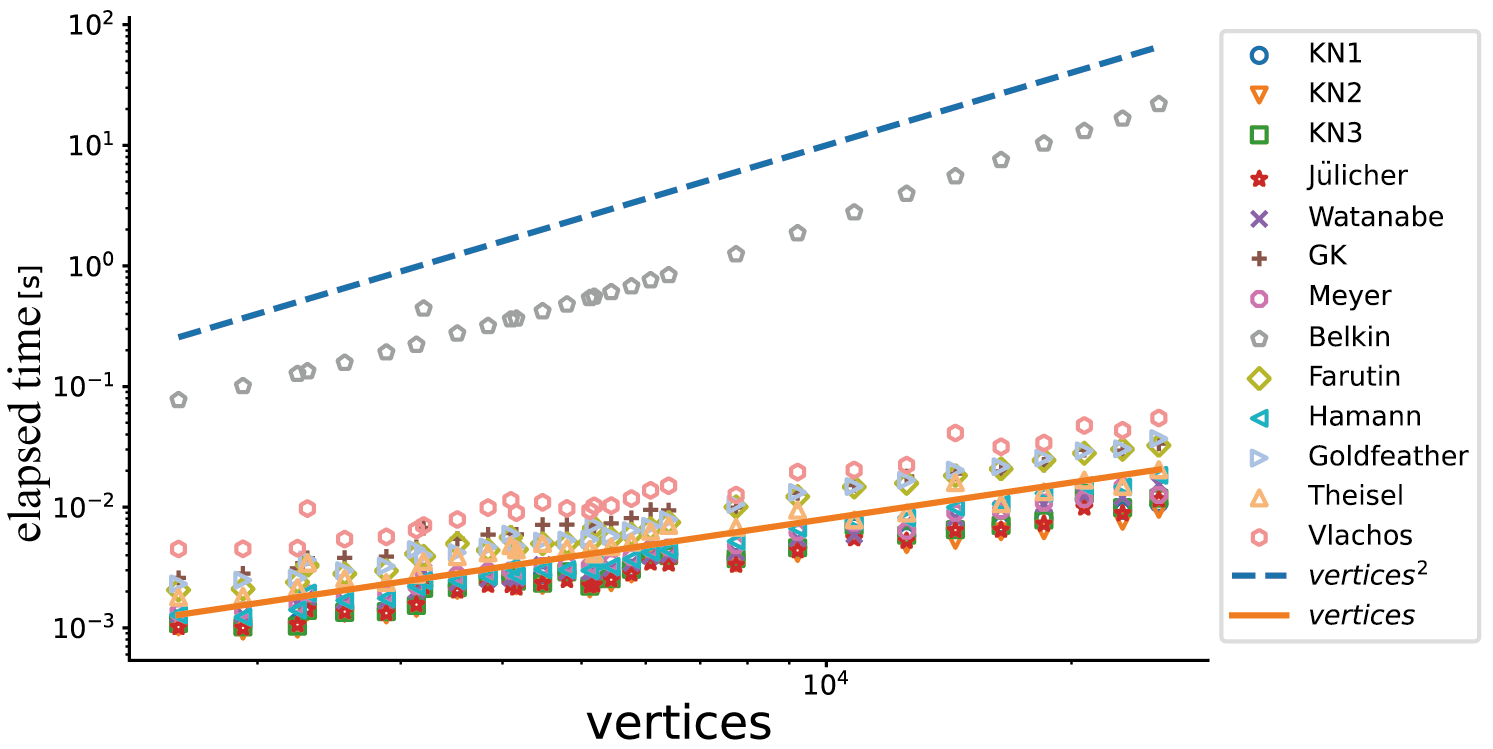}
    \caption{Relationship between the number of vertices and the computation time for each bending model in the calculation of Fig. \ref{Fig.3-2}. The blue dotted line is the straight line for $O(N^2)$ scaling, and the orange solid line is the straight line for $O(N)$ scaling.}
    \label{Fig.3-7}
\end{figure}

Among the models using triangular dihedral angles, the energies of the KN2 and KN3 models deviated from the theoretical values owing to the equilateral triangle approximation, which resulted in an anisotropic mesh when the cells were packed in one direction, as shown in Figs. \ref{Fig.3-2}(a) and (b). The energy error of the KN1 model is comparable to those of the GK and Meyer models for the (b)$y$-directional division; however, it exhibits a significant error for the (a)$x$-directional division. This is because the bending in Fig. \ref{Fig.3-2}(b) occurs along each edge, whereas the bending and edge directions are not aligned in Fig. \ref{Fig.3-2}(a), as evident from the cell arrangements shown on the left-hand side in Fig. \ref{Fig.3-1}. For the structured mesh, the error of the KN1 model was smaller than those of the KN2 and KN3 models (Fig. \ref{Fig.3-2}). However, for the unstructured mesh as shown in Figs. \ref{Fig.3-3}(b) and \ref{Fig.3-4}(b) show that the errors of the KN1 model are worse than those of the KN2 and KN3 models, which are equilateral triangle approximations. Fig. \ref{Fig.3-4}(a) further illustrates that the KN1 model has a larger error than the other discrete curvature models. This indicates that the weights of the KN1 model $l_{ij}/\sqrt{3}\sigma_{ij}$ are ineffective in many cases. 

Because the Belkin model is inherently designed for spherical meshes, it exhibits significantly smaller energy and force-density errors, as shown in Fig. \ref{Fig.3-4}. However, the errors are substantially larger for planar meshes (Figs. \ref{Fig.3-2} and \ref{Fig.3-3}), which can be attributed to the dependence of the parameter $h$ in Eqs. (\ref{Eq.Belkin1}) and (\ref{Eq.Belkin2}) on the mesh shape\cite{Li2015}. Consequently, the Belkin model is not suitable for scenarios with dynamically changing mesh structures, such as the current cell-division simulations, because the optimal parameter $h$ can also vary. 

The J\"{u}licher, GK, and Meyer models exhibit comparable errors. However, the GK model demonstrates a notably large error in the force-density calculation, as shown in Fig. \ref{Fig.3-4}(b). This arises because $A_i^{GK}$ can become zero or negative for anisotropic meshes owing to the inclusion of a cotangent in the expression of $A_i^{GK}$ in Eq. (\ref{Eq.GK2}). Therefore, the GK model is not considered suitable for the cell-center model because of its large error. As shown in Fig. \ref{Fig.3-7}, the computational costs of the J\"{u}licher and Meyer models is comparable to that of the two-sided angle model. Figs. \ref{Fig.3-3}(b) and \ref{Fig.3-4} indicate that the Meyer model has a smaller force-density error for unstructured meshes. By contrast, the J\"{u}licher model does not have any cotangent expressions.

Unlike the other models, the Watanabe model exhibited a large error for the structured mesh. We speculated that this is because the expression for the normal curvature in Eq. (\ref{Eq.Watanabe2}) underestimates the curvature value low, as $\kappa^{ij}_n$ cancels out between opposite edges of regularly arranged cells. 

The Farutin, Hamann, and Goldfeather models, which employ the least-squares method, exhibited no significant differences in accuracy. The Goldfeather model exhibits exceptionally high accuracy when the normal vector is known \cite{Goldfeather2004,Zhihong2011}. However, its accuracy diminishes when the estimated normal vectors are used, as in this study. Moreover, because there is no significant error differences between the Farutin model, which uses the values of the partial derivatives, and the Hamann model, which uses the coefficients of the fundamental form of surfaces, the computationally efficient Hamann model is considered sufficient.

Finally, the results of the Theisel and Vlachos models, which employ triangle interpolation, were similar to those obtained using the least-squares method because estimated normal vectors were used. However, as shown in Fig. \ref{Fig.3-4}, the Vlachos model exhibits a larger error than the other models for spherical surfaces. This discrepancy likely arises from the inherent characteristics of B\'ezier surfaces and their implications for curvature calculations. The high flexibility of B\'ezier surfaces, while advantageous for representing diverse shapes, can lead to challenges in approximating surfaces that are not perfectly smooth. We believe that there is room for improvement in the Vlachos model as it uses a B\'ezier curve and the selection of control points can significantly affect the results. Because the computational cost of the Theisel model is slightly higher than that of the Hamann model, Hamann is sufficient.

\subsection{Dynamic condition}\label{sec4-2}
Fig. \ref{Fig.3-6} reveals two major patterns for $K_Z\ge1$. The models clustered on the lower side exhibit large force-density errors, as shown in Fig. \ref{Fig.3-3}(b), and are considered to be in a quasi-stable state. By contrast, the slope of the models on the upper side is closer to the theoretical value of 0.25.

For the $x$-directional division, the wavenumbers of the KN2 and KN3 models are larger than those of the other models from $K_Z=10^{-1}$ to $1$. This can be attributed to the fact that the weighting factor $l_{ij}/\sqrt{3}\sigma_{ij}$ of the edge related to bending in the $x$-direction was considered to be greater than 1. Consequently, the unweighted KN2 model and KN3 model estimate a lower bending stiffness in the $x$-direction, rendering the sheet softer and resulting in a larger wavenumber.

\subsection{Limitation}\label{sec4-3}
While the parameter values used in this study (as listed in Table 1) are based on previous research \cite{MIMURA2023} and allow for reproducible simulations in a 3D vertex model, they do not represent the exact parameters of biological tissues. Real tissues exhibit more complex biomechanical properties, including the influence of basement membranes, varying mechanical constraints, and heterogeneous material properties. These factors are not fully accounted for in the our model. 

This study employed a cell-center model to validate simulations using a specific parameter set and cell numbers. Consequently, the consistency of results cannot be guaranteed under varying conditions, such as different parameter sets or cell numbers. However, in the static analysis (Section \ref{sec3-1}), we evaluated the error related solely to bending energy by averaging results across meshes with different vertex counts. This approach suggests the observed trends are robust within similar setups. Still, future research is needed to investigate the results' robustness across a wider range of conditions, further validating the generality of the findings.

\section{Conclusion}\label{sec5}
In this study, we compared 13 bending-energy models in a center-cell model, wherein anisotropic meshes are generated during computation. To assess the accuracies of these models, we first prepared sheets with various cell arrangements under a simple wrinkle pattern for which the theoretical bending energy could be computed and compared them with the bending energies computed by each model. Next, we examined the wrinkle and folding structures of the planar cell sheets when they were divided by specifying the cell-division axes in the uniaxial and random directions, respectively. 

Among the models evaluated for handling bending deformations caused by cell division in the cell-center model, the Hamann, Goldfeather, and Farutin models, which employ the least-squares method, exhibited considerably small errors. Therefore, these models are suitable for handling bending deformations in the cell-center model of morphogenesis. Furthermore, because the accuracies of these models did not differ significantly, the Hamann model is recommended as it inccurs the lowest computational cost.

However, all models consistently predicted that wrinkles become finer as the out-of-plane deformation constraint energy coefficient $K_Z$ increases. Therefore, as far as qualitative properties are concerned, the J\"{u}licher model, which is more computationally less expensive, can be used.

\section*{CRediT authorship contribution statement}
Tomohiro Mimura: Conceptualization, Methodology, Software, Investigation, Visualization, Writing-original draft, Funding acquisition.Yasuhiro Inoue: Conceptualization, Methodology, Supervision, Writing-review \& editing, Funding acquisition.

\section*{Declaration of competing interest}
The authors declare that they have no known competing financial interests or personal relationships that could have appeared to influence the work reported in this paper.

\section*{Acknowledgement}
This study was supported by MEXT KAKENHI [grant number 20H05947] and JST SPRING [grant Number JPMJSP2110]. We would like to thank Editage (www.editage.jp) for English language editing.

\section*{Data availability}
The data associated with this study are available at:\\  https://github.com/TomohiroMimura/BendingEnergyDiscretization.

\begin{table}[h]
    \caption{Parameters used for out-of-plane deformation simulations}
    \label{table1}
    \centering
     \begin{tabular}{clll}
      \hline
      Symbol & Value & Description \\
      \hline \hline
      $\eta$ & 0.25 & {Cell-point friction coefficient } & \\
      $K_L$ & 2 & {Edge-length elasticity constant} & \\
      $K_S$ & 10 & {Triangular area elastic constant} \\
      $\tilde{K_B}$ & 10 & {Dihedral-angle elastic constant} \\
      $K_Z$ & $10^{-3}$-4 & {Out-of-plane deformation constraint constant}\\
      $K_{\rm{Rep}}$ & 10 & {Cell-repulsion constant} \\
      $K_{\rm{Center}}$ & 10 & Constraint constant \\
      $L_{\rm{eq}}$ & 1.0 & {Equilibrium-edge length} \\
      $S_{\rm{eq}}$ & 0.433 & {Equilibrium-triangle area} \\
      $\theta_0$ & 3.14159 & Dihedral-angle equilibrium\\
      $H_0$ & 0 & Mean Curvature equilibrium\\
      $r_c$ & 0.9 & {Cut-off distance}\\
      $l_{\rm{th}}$ & 0.25 & Flip-threshold length\\
      \hline
      $\Delta t$ &0.0002 & Integration time step\\
      $\tau_{i}^{\rm{cycle}}$ & 100 &Cell cycle\\
      {$\tau_{\rm{ct}}$} & {0.001} & {Flip cool time}\\
      \hline
      \\
     \end{tabular}
\end{table}






\clearpage

\bibliographystyle{elsarticle-num} 
\bibliography{main}

\end{document}